\documentclass[a4paper,fleqn]{cas-dc}

\usepackage[numbers]{natbib}
\usepackage{lineno,hyperref}
\usepackage{comment}
\usepackage{framed}
\usepackage{xspace}
\usepackage{breakcites}
\usepackage[framemethod=tikz]{mdframed}
\definecolor{shadecolor}{rgb}{0.75, 0.75, 0.75}
\definecolor{light-gray}{gray}{0.95}
\usepackage{float}
\usepackage{booktabs}
\usepackage{lscape}
\usepackage{longtable}
\usepackage{soul}
\usepackage{fixltx2e}
\usepackage[utf8]{inputenc}
\usepackage[T1]{fontenc}
\usepackage{ulem}
\usepackage{multibib}
\usepackage{breakcites}
\usepackage{caption}
\newcites{sec}{Secondary Studies}
\newcites{secNot}{Secondary Studies}

\newcommand{\MyBox}[1]{\vspace{3mm}\noindent\framebox[\columnwidth][c]{\parbox[b]{0.95\columnwidth}{ #1 }}\vspace{3mm}}

\begin{document}
\let\WriteBookmarks\relax
\def\floatpagepagefraction{1}
\def\textpagefraction{.001}
\shorttitle{Grey Literature in Software Engineering: A Critical Review}
\shortauthors{Kamei et al.}

\title [mode = title]{Grey Literature in Software Engineering: A Critical Review}                      



\author[1,2]{Fernando Kamei}[
                        orcid=0000-0002-5572-2049]
\cormark[1]
\ead{fernando.kenji@ifal.edu.br}

\author[3]{Igor Wiese}[]
\author[4]{Crescencio Lima}[]
\author[3]{Ivanilton Polato}[]
\author[5]{Vilmar Nepomuceno}[]
\author[6]{Waldemar Ferreira}[]
\author[7]{M\'arcio Ribeiro}[]
\author[2]{Carolline Pena}[]
\author[8]{Bruno Cartaxo}[]
\author[9]{Gustavo Pinto}[]
\author[2]{S\'ergio Soares}[]

\address[1]{Federal University of Pernambuco, Recife, Pernambuco, Brazil}
\address[2]{Federal Institute of Alagoas, Macei\'o, Alagoas, Brazil}
\address[3]{Federal University of Technology of Paran\'a, Campo Mour\~ao, Paran\'a, Brazil}
\address[4]{Federal Institute of Bahia, Vit\'oria da Conquista, Bahia, Brazil}
\address[5]{Federal Institute of Pernambuco, Recife, Pernambuco, Brazil}
\address[6]{Universidade Cat\'olica de Pernambuco, Recife, Pernambuco, Brazil}
\address[7]{Federal University of Alagoas, Macei\'o, Alagoas, Brazil}
\address[8]{Federal Institute of Pernambuco, Paulista, Pernambuco, Brazil}
\address[9]{Federal University of Par\'a, Bel\'em, Par\'a, Brazil}

\cortext[cor1]{Corresponding author}


\begin{abstract}
Context: Grey Literature (GL) recently has grown in Software Engineering (SE) research since the increased use of online communication channels by software engineers. However, there is still a limited understanding of how SE research is taking advantage of GL.\\
Objective: This research aimed to understand how SE researchers use GL in their secondary studies.\\
Method: We conducted a tertiary study of studies published between 2011 and 2018 in high-quality software engineering conferences and journals. We then applied qualitative and quantitative analysis to investigate 446 potential studies.\\
Results: From the 446 selected studies, 126 studies cited GL but only 95 of those used GL to answer a specific research question representing almost 21\% of all the 446 secondary studies. Interestingly, we identified that few studies employed specific search mechanisms and used additional criteria for assessing GL. Moreover, by the time we conducted this research, 49\% of the GL URLs are not working anymore. Based on our findings, we discuss some challenges in using GL and potential mitigation plans.\\
Conclusion: In this paper, we summarized the last 10 years of software engineering research that uses GL, showing that GL has been essential for bringing practical new perspectives that are scarce in traditional literature. By drawing the current landscape of use, we also raise some awareness of related challenges (and strategies to deal with them). 
\end{abstract}


\maketitle

\section{Introduction}
Grey Literature (GL) is considered to be a kind of literature that has not been subject to quality control mechanisms
(e.g., the peer reviewed process) before publication~\cite{Petticrew:2006:Book:SR}.
Over recent years, GL stands out as an essential source of knowledge to be used alone or complementing research findings within the traditional literature~\cite{Garousi:2016:EASE}. Diversity of scientific areas, e.g., in medicine~\cite{Paez:2017:Medicine}, in management~\cite{Adams:2017:Shades:IJMR}, and nutrition~\cite{Adams:2016:Health}, have investigated the GL. 
Some researchers observed that benefits of using GL included gaining significant knowledge from practitioners in addition to academic articles~\cite{Adams:2017:Shades:IJMR}, reducing publication bias where studies only report the positive findings to be published~\cite{Paez:2017:Medicine}, and in a way to address topics that are missing from conventional academic sources~\cite{Sumeer:2020}.

The interest of GL in the context of Software Engineering (SE) is more recent. It has increased over the last few years, mainly due to the widespread presence of GL media used by SE professionals, including various types of social media and publication channels~\cite{Storey:2014:ESM}. For Rainer~\cite{Rainer:2017:IST}, those media, in their distinct nature, that practitioners are producing could help researchers gain additional insights into the dynamics and challenges that occur during the software development process. This potentially explains why SE researchers have been paying attention to the potential of GL lately. Several works explore, for example, its relationship to question and answer (Q\&A) websites such as Stack Overflow~\cite{Zahedi:2020:EASE}; and news aggregator websites, such as Reddit and Hacker News~\cite{Aniche:ICSE;2018}.

\begin{figure}[h!]
\centering
\includegraphics[scale = 0.3, clip = true, trim= 80px 50px 50px 50px]{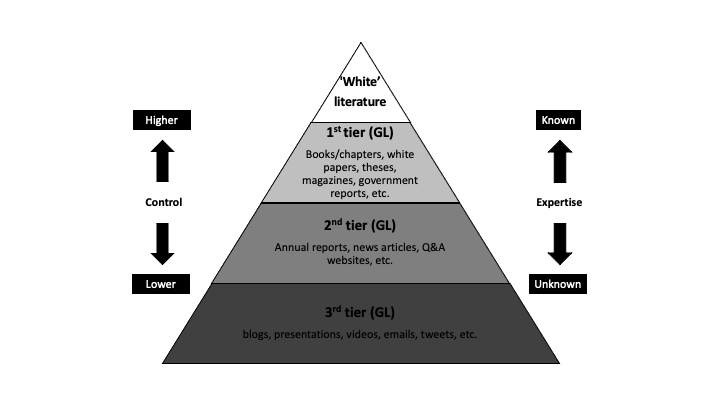}
\caption{The ``shades'' of Grey Literature for Software Engineering, adapted from Garousi et al.~\cite{Garousi:2019:IST}.}
\label{fig:shades-gl}
\end{figure}

Adams et al.~\cite{Adams:2016:Health} introduce the idea of ``grey information'' to distinguish different grey forms, including grey literature, grey information, and grey data. The term ``grey data'' is used to describe user-generated web content in SE. For instance, Williams and Rainer's study~\cite{Williams:2019:EASE} considered grey data the content of tweets, blogs, and posts on Q\&A websites. On the other hand, ``grey information'' is informally published or not published at all, e.g., meeting notes and emails. These detailed categories have not been widely adopted in the SE literature~\cite{Rainer:2018:TR}. Similarly, we considered all forms of grey data and grey information as GL in our work. Due to the diversity in GL types, Adams et al.~\cite{Adams:2017:Shades:IJMR} classified them according to ``shades'' of grey. The same position was adopted in SE research by Garousi et al.~\cite{Garousi:2019:IST}, adapting these shades according to three-tiers (see Figure~\ref{fig:shades-gl}). On the top of the pyramid is the White Literature (i.e., published journal papers or conferences, proceedings). The other tiers are composed of the GL sources classified according to the shades of grey. These tiers are arranged according to two dimensions: expertise and outlet~\cite{Garousi:2020:book}. The first one runs between extremes ``unknown'' and ``known,'' and the second one runs between the extremes ``lower'' and ``higher.'' The darker the color, the less moderated or edited is the source in conformity with explicit and transparent knowledge creation criteria. In our work, we adopted the ``shades'' of grey as a reference to guide our interpretation of GL types throughout the research.

Along with GL's growth as a scientific area of study, SE researchers started to expand their views on the use of GL through the lenses of secondary studies. For instance, Garousi et al.~\cite{Garousi:2019:IST} proposed a Multivocal Literature Review (MLR) method to incorporate GL alongside traditional literature. Similarly, Raulamo-Jurvanen et al.~\cite{Raulamo-Jurvanen:2017:EASE} conducted a Grey Literature Review (GLR) to understand how software practitioners choose the right test automation tool. 
There are also invaluable tertiary studies that gather a broader perception of GL usage~\cite{Neto:2019:ESEM,Yasin:Thesis:2020,Zhang:2020:ICSE}.

This paper aims to contribute to the literature by compiling additional evidence on the use, motivations, benefits, and challenges of using GL in Secondary Studies. Our work is unique because we explored the criteria employed in Secondary Studies to select studies, the coverage of GL to support answers to research questions, the motivations of studies in their decision to use or to avoid GL, and the definitions of GL used in different circumstances. To achieve this goal, we performed a tertiary study using automatic and manual searches during 2011 and 2018. 
Our search process identified 20,181 studies, from which we identified 446 Secondary studies that fulfilled our eligibility criteria. We noticed that 126 out of 446 Secondary Studies used or searched for GL. We used these two sets (446 and 126 Secondary Studies) for a more in-depth analysis, using qualitative and quantitative approaches. Our main findings are the following:

\begin{itemize}
    \item GL is not extensively used in Secondary Studies of SE (126/446), although we noticed a growth over the years;
    \item At least 75\% of the studies (95/126) used GL to support answers to at least one research question, even GL represents less than 21\% of all the 446 Secondary Studies;
    \item Misunderstandings about GL types were sometimes controversial elements among the studies;
    \item Almost 50\% of the GL reported in investigated studies (n=126) are now unavailable;
    \item Few studies (14/126) employed specific criteria to search for and additional criteria for assessing the GL quality (7/126);
    \item Consultants and companies were the ones that most produced the GL found. Although, there is an increasing amount of content produced by practitioners over the years;
    \item Diverse challenges in dealing with GL were found that SE researchers may have to face. Nevertheless, we provided a potential list of ways to address that could help SE researchers to deal with each one.
\end{itemize}

By describing these findings and a list of challenges with potential ways to deal with them for SE researchers, we expect to help others better conduct secondary studies using GL to take advantage of SE practice. 

We organized the rest of the work as follows: Section~\ref{sec:related_works} introduces the related works compared to this study. Section~\ref{sec:RQ} poses our research questions. Section~\ref{sec:method} presents the methods we employed to conduct a tertiary study. Section~\ref{sec:results} provides answers to each research question. Section~\ref{sec:discussion} discusses some of our findings and the implications of this research. Section~\ref{sec:challenges-recommendations} shows challenges for dealing with GL in SE research with some potential ways to address them. Section~\ref{sec:limitations} discusses some threats to validity. Finally, Section~\ref{sec:conclusions} concludes our work and presents directions for future works.

\section{Related Work}\label{sec:related_works}

Software Engineering (SE) studies that target Grey Literature (GL) are particularly recent (e.g., ~\cite{Garousi:2016:EASE,Williams:2018:EASE}).  
GL is not only used in primary studies but also in secondary ones (e.g.,~\cite{ Garousi:2016:EASE,Raulamo-Jurvanen:2017:EASE}) and tertiary studies (e.g.,~\cite{Neto:2019:ESEM,Zhang:2020:ICSE,Yasin:Thesis:2020}). In this section, we focused on GL used in secondary and tertiary studies.

Garousi et al.~\cite{Garousi:2016:EASE} were one of the first to investigate the use of GL, in addition to traditional literature, in Secondary Studies. The authors compared the results of two investigations: one included GL while the other did not. Their findings highlighted the importance of using GL to cover technical research questions. Garousi and colleagues~\cite{garousi2017guidelines,Garousi:2019:IST} continued their investigations on using GL proposed the use of MLR. Since then, the SE community has conducted some MLR studies (e.g.,~\cite{Garousi:2018:JSS,Tripathi:2018:MLR}). 

The work of Raulamo-Jurvanen et al.~\cite{Raulamo-Jurvanen:2017:EASE} conducted the first GLR we have learned about in SE, aiming to understand how software practitioners choose the right test automation tool. Their findings are mostly derived from practitioners' experiences and opinions that work for consulting companies or tool vendors. Aiming to improve the credibility of their findings, they employed some criteria to assess the evidence of the GL, including, for instance, the number of readers, the number of comments, or the number of hits on Google.

There are also tertiary studies that explored the use of GL in secondary studies~\cite{Neto:2019:ESEM,Yasin:Thesis:2020,Zhang:2020:ICSE}. Yasin et al.~\cite{Yasin:Thesis:2020} investigated the evidence of GL use in secondary studies published until 2012. Their investigations found GL in 76\% of the Secondary Studies and that the level for GL evidence of Systematic Literature Review (SLR) synthesis discussion was around 9\%. This work employed some GL definitions; for instance, it mentioned the Luxembourg definition. It was then mentioned that GL is always referred to as ``fugitive literature'' as it is semi-published. This study considered theses, conference proceedings, technical reports, official documents, company white papers, discussion boards, and blogs as GL types.
\begin{itemize}
    \item \textit{Differences from our work:} in short, our study expands this investigation to cover the studies published until 2018. Moreover, we expanded our investigation's scope (we explored the methods of studies used to collect and to assess GL's quality, motivations for use and reasons to avoid it, and perceived benefits and challenges of its use). The main distinctions made are for the period investigated, the coverage of sources to search for studies, and different interpretations about GL types. This work is particularly interesting to provide the first overview about GL in Secondary Studies in SE.
\end{itemize}

Neto et al.~\cite{Neto:2019:ESEM} conducted a tertiary study that focused only on MLR and GLR, aiming to provide a preliminary investigation about research involving these types of studies. The research focused on understanding their (i) motivations to included GL (e.g., lack of academic research on the topic and evidence in GL), (ii) the types of GL used (e.g., videos, books, blog post, and technical report), and (iii) the search engines used (e.g., Google, Google Scholar, and websites). They searched for the studies published between 2009 and April 2019, using six academic search engines. The search returned fifty-six studies. Twelve of them were selected. 

\begin{itemize}
    \item \textit{Differences from our work:} like our study, Neto and colleagues also investigated secondary studies; however, the main difference between our research and theirs is that we investigated any kind of Secondary Study to gain an overview of GL in Secondary Studies. Beyond that, we also investigated the studies' motivations that did not use GL in research and in-depth investigation of GL use between the Secondary Studies.
\end{itemize}

The recent study by Zhang et al.~\cite{Zhang:2020:ICSE} is the closest to our work to the best of our knowledge. Zhang et al.'s study investigated GL using a mixed-methods approach by conducting a tertiary study and a survey with SE researchers that used GL in Secondary Studies. They aimed to obtain an overview of the research community's understanding of (i) the possible definitions of GL in SE (they did not find a standard definition), (ii) the reasons for including GL from the perspectives of both literature users and community experts (e.g., to seek more related research and to avoid publication bias), (iii) proposing a conceptual model for how SE researchers work with GL in the research life-cycle, and (iv) identifying the significant challenges of GL use in SE (e.g., lack of understanding of GL and difficulty in quality assessment). 
\begin{itemize}
    \item \textit{Differences from our work:} our work differs in at least three dimensions: (i) we conducted a broader search to retrieve the highest number of studies published on premier SE conferences and journals; (ii) we explored waters not chartered by previous studies (e.g., we investigated the inclusion/exclusion criteria, the perspective on GL use according to the types of Secondary Studies, and the availability of the GL data, investigating the extent to which the research questions are answered using GL); and (iii) we found some different findings that need further investigation. For instance, Zhang et al.~\cite{Zhang:2020:ICSE} found that only 25\% of researchers used GL to evaluate their conclusions, and we found that more than 50\% of the secondary studies used GL to support their findings. Moreover, the definitions of GL found in our work were slightly different.
\end{itemize}

\section{Research Questions}
\label{sec:RQ}
The general \textbf{goal} of this research is to understand how Software Engineering (SE) researchers use and take advantage of GL in their secondary studies.

This tertiary study is motivated by the following six research questions (RQs):

\MyBox{\textbf{RQ1. \textit{What definitions of Grey Literature are employed in Secondary Studies?}}}

\noindent
\textit{Rationale:} Since the GL is recent in SE research, the work of Zhang et al.~\cite{Zhang:2020:ICSE} did not find a common definition for GL in SE research that might influence its use. According to Bonato~\cite{Bonato:SearchingGLBook:2019}, it is essential to define GL as a recent area that will make it easier to search for and assess a GL source. This research question intends to investigate the way GL is defined. Answering this question is essential to improve the state of the art of GL in SE.

\MyBox{\textbf{RQ2. \textit{How is Grey Literature used in Secondary Studies?}}}
\noindent
\textit{Rationale:} When conducting secondary studies, researchers often employ inclusion criteria to filter out not peer reviewed works. This decision is motivated by the fact that peer reviewed work is considered more reliable than not peer reviewed work~\cite{Dina:2014:SLR}. However, more recently, a growing number of researchers are arguing that not peer reviewed work could also be incorporated into scientific studies~\cite{Garousi:2016:EASE}. Some studies even go further and consider not peer reviewed works as the exclusive data source~\cite{Soldani:2018:GLR}. In this research question, we sought to understand if secondary studies are using GL and, if so, how. More precisely, we aim to investigate (i) the frequency of studies using GL; (ii) the frequency of GL use between the types of Secondary Studies that used it; and (iii) the frequency to which GL is used to support answers to research questions.

\MyBox{\textbf{RQ3. \textit{How is Grey Literature searched, selected, and has the quality assessed in Secondary Studies?}}} \noindent
\textit{Rationale:} Using or searching for GL is not a trivial task. Some of the reasons for this lie in the diversity of its sources. GL differs in the type of structure and content provided by each source. This could even make it difficult, for instance, to search for or find information in GL~\cite{Kamei:SBES:2020}. In this research question, we sought to investigate the methods employed to (i) search, (ii) select, and (iii) perform a GL quality assessment. A better understanding of these procedures would be essential to guide future research in this area.

\MyBox{\textbf{RQ4. \textit{What types of Grey Literature are the most frequently used in Secondary Studies?}}}

\noindent
\textit{Rationale:} GL is available in many forms, varying from traditional mediums, such as books and technical reports, to more dynamic mediums, such as forums and Question \& Answer websites. These mediums offer researchers a rich spectrum of unstructured data, which brings specific benefits and limitations. However, since these mediums are often hosted on specific platforms (e.g., from official websites to personal blogs), it is not clear if (and how long) this information will be accessible. In this research question, we sought to investigate more concretely (i) the different types of GL used; (ii) who are the producers; and (iii) the availability of the GL. A better understanding of the GL source would be important to guide future research in this area.

\MyBox{\textbf{RQ5. \textit{What motivates researchers to use/avoid Grey Literature?}}}

\noindent
\textit{Rationale:} The process of conducting a literature review is far from trivial. It is important to plan according to the research focus.
In this question, we want to understand if researchers are properly justifying GL's use in their research. More concretely, we aim to investigate (i) the motivations to use and (ii) the reasons to avoid GL.  We know aware that this information is sometimes not explicitly described in the research papers but rather implied in the context. Answering this question is important for SE researchers who do not know how GL could improve their research.

\MyBox{\textbf{RQ6. \textit{How do researchers perceive the use of Grey Literature?}}}

\noindent
\textit{Rationale:} Some researchers advocate for GL because of some known benefits, including practical orientation and the appeal to complement formal literature. In this research question, we intend to complement the (i) perceived benefits of using GL, and wait to uncover eventually (ii) challenges related to its usage. Answers to this question could support SE researchers in understanding potential benefits that could be unlocked (and the challenges behind them).

\section{Research Method}\label{sec:method}

In this tertiary study, we followed the guidelines of Kitchenham et al.~\cite{Kitchenham:2015:ESE:BOOK}. In what follows, we present the procedures employed to search for the studies and the steps taken for selection, data extraction, and data analysis.

\subsection{Search strategy}\label{sec:phases1-4}

We began the search procedures for Secondary Studies in 2019, using a combination of three approaches: (i) a selection of previous tertiary studies, (ii) automatic searches in digital libraries, and (iii) a manual analysis of a small selection of premier SE journals and conference proceedings. 

Since we were not interested in specific studies (e.g., a tertiary study about pair programming), we started by searching for tertiary studies that focused on general aspects of SE. We selected the following tertiary studies: Kitchenham et al.~\cite{Kitchenham:2009:IST}, Kitchenham et al.~\cite{Kitchenham:2010:IST}, Da Silva et al.~\cite{daSilva:2011:IST}, Cruzes and Dyb{\aa}~\cite{Cruzes:2011:IST}, and Neto et al.~\cite{Neto:2019:ESEM}. These studies covered Secondary Studies published until 2010, excluding the Neto et al.'s study~\cite{Neto:2019:ESEM}, which covered only MLR and GLR studies until 2019. We complement the search with an automatic and a manual search to find studies published between 2011 and 2018.

We used the most relevant SE digital libraries for the automatic search, as recommended by Kitchenham et al.~\cite{Kitchenham:2015:ESE:BOOK}, namely ACM Digital Library, IEEE Xplore, ScienceDirect, and Scopus.

For the manual search, we chose the most prestigious SE journals and conferences related to this research topic, namely: 

\begin{itemize}
	\item Journals: \textit{ACM Transactions on Software Engineering Methodology (TOSEM), IEEE Transactions on Software Engineering (TSE), Empirical Software Engineering Journal (EMSE), Information and Software Technology (IST)}, and \textit{Journal of Systems and Software (JSS);}
	\item Conferences: \textit{International Conference on Software Engineering (ICSE), Empirical Software Engineering and Measurement (ESEM)}, and \textit{Evaluation and Assessment in Software Engineering (EASE).}
\end{itemize}

\subsubsection{Search string}
\label{sec:method-searchterms}

We took advantage of the same search string adopted by Da Silva et al.~\cite{daSilva:2011:IST} to conduct our search. In our case, we adapted this search string to cover additional terms, such as Grey Literature Reviews and Multivocal reviews. The updated search string is as follows: \\
\framebox{
	\parbox[t][5.2cm]{8cm}{
		\addvspace{0.1cm} \centering 
		\textit{(``software engineering'') AND (``review of studies'' OR ``structured review'' OR ``systematic review'' OR ``grey review'' OR ``grey literature'' OR ``gray review'' OR ``gray literature'' OR ``multivocal literature'' OR ``multi-vocal literature'' OR ``literature review'' OR ``literature analysis'' OR ``in-depth survey'' OR ``literature survey'' OR ``meta analysis'' OR ``past studies'' OR ``subject matter expert'' OR ``analysis of research'' OR ``empirical body of knowledge'' OR ``overview of existing research'' OR ``body of published research'' OR ``evidence-based'' OR ``evidence based'' OR ``study synthesis'' OR ``study aggregation'')}
	} 
}

\vspace{0.3cm}
The Scopus library has a limited length on its search field. In this case, we had to break down the search string into 24 sub-queries. 
After we retrieved all studies from the search procedures, we organized them all onto a shared spreadsheet, which we used to conduct the next methodological steps.

\subsection{Selection criteria}
\label{sec:selecting-study}

When manually investigating the retrieved papers, we focused on selecting Secondary Studies. We paid particular attention to the following kinds of secondary studies:

\begin{itemize}
	\item Systematic Literature Review (SLR) (or Systematic Review);
	\item Mapping Study (MS) (or Systematic Mapping Study);
	\item Meta-Analysis (MA);
	\item Grey Literature Review (GLR);
	\item Multivocal Literature Review (MLR).
\end{itemize}

For each candidate paper, we applied a set of \textbf{exclusion} criteria. Table~\ref{tab:ec} describes each exclusion criterion. We excluded any candidate study that complies with at least one exclusion criterion. The only exception is the criterion EC1, which was not applied to the studies retrieved from previous tertiary studies.

\begin{table}[!ht]
\caption{List of exclusion criteria.}
\small
\centering
\begin{tabular}{cp{7cm}}
\toprule
\textbf{\#} & Description \\
\midrule
EC1 & The study was published before 2011 or after 2018.\\
EC2 & The study was duplicated.\\
EC3 & The study was not written in English.\\
EC4 & The study was not a full paper (e.g., position papers, abstracts, posters, etc).\\
EC5 & The study was not peer reviewed (e.g., editorials, summaries, letters, keynotes, slides, etc).\\
EC6 & The study did not report a Secondary Study (i.e., SLR, MS, MLR, GLR, or MA).\\
EC7 & The study was not related to Software Engineering (e.g., Information Systems and Computer Science).\\
EC8 & The venue in which the Secondary Study was published did not have a minimum h5-index (20 for conferences and 25 for journals).\\
\bottomrule
\end{tabular}%
\label{tab:ec}
\end{table}

\subsection{Study selection}

\begin{figure*}
	\centering
	\includegraphics[scale = 0.5, clip = true, trim= 0px 0px 0px 0px]{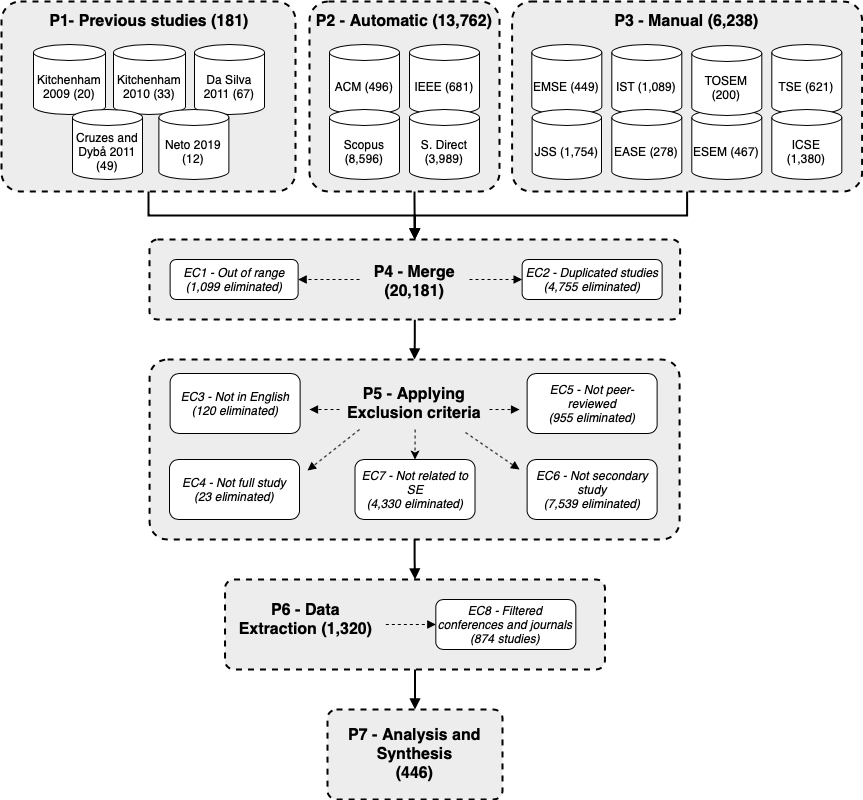}
	\caption{Process of selecting studies in each phase of the tertiary study.}
	\label{fig:selection-process}
\end{figure*}

The study selection procedure was conducted in seven sub-phases, as depicted in Figure~\ref{fig:selection-process}. There is a number indicating each phase (\textbf{P1--P7}).

At phase \textbf{P1}, we selected a total of 181 Secondary Studies cited in the tertiary studies. We found these studies at Kitchenham et al. (2009)~\cite{Kitchenham:2009:IST} (20 studies), Kitchenham et al. (2010)~\cite{Kitchenham:2010:IST} (33 studies), Da Silva et al.~\cite{daSilva:2011:IST} (67 studies), Cruzes and Dyb{\aa}~\cite{Cruzes:2011:IST} (49 studies), and Neto et al.~\cite{Neto:2019:ESEM} (12 studies).

At phase \textbf{P2}, we selected a total of 13,762 studies using the automated search. We found these studies at ACM Digital Library (496), IEEE Xplore (681), Science Direct (3,989), and Scopus (8,596). 

At phase \textbf{P3}, we found 6,238 studies using the manual search. We found these studies at EMSE (449), IST (1,089), TOSEM (200), TSE (621), JSS (1,754), EASE (278), ESEM (467), and ICSE (1,380). The phases P1--P3 retrieved a total of 20,181 potential studies. Each potential study retrieved received a singular identification (ID).

Next, at phase \textbf{P4}, we sorted the Secondary Studies by title and organized them on a spreadsheet. We applied the EC1 and EC2 to remove the studies out of the range of our investigation and the studies with the same bibliographical information (i.e., title, abstract, and author(s)). For EC2 criterion, we employed the following next steps: (i) We compared paper titles; (ii) For papers with the same title, we looked at the abstracts and; if they are different, we considered the complete study as recommended by Kitchenham and Charters~\cite{Kitchenham:2007:Guideline}; if they are the same, we exclude one of them, if the publication years are different, we excluded the least recent study. We removed 5,854 studies, 1,099 studies published before 2011 or after 2018 (EC1), and 4,755 instances of duplicated studies (EC2), respectively. At the end of this phase, 14,327 studies remained.

At phase \textbf{P5}, we read the studies thoroughly and applied the exclusion criteria (EC3--EC7) to all the 14,327 potentially relevant studies. 

Determining whether a Secondary Study fits in the inclusion/exclusion criteria is a subjective task. To reduce the subjectivity and gain an alignment of understanding among the researchers involved in this work, we applied a pilot of the criteria to a small set of the selected studies. 
Since it would be too time-consuming to apply the exclusion criteria in pairs for all selected works, we applied those criteria in pairs in a random sample of 21\% (=3,030) of the total of studies. Six authors participated in this review process (the first author paired with the additional co-authors).
Each author applied the criteria individually, and, in the case of disagreements, we discussed them in conflict resolution meetings. In the case that no agreement was achieved, a third author joined the discussion. 
To evaluate the agreement level, we performed an agreement analysis using the Kappa scale~\cite{Landis:1977:Kappa}. Kappa scores are generally interpreted as slight ($\geq$ 0 and $\leq$ 0.20), fair ($\geq$ 0.21 and $\leq$ 0.40), moderate ($\geq$ 0.41 and  $\leq$ 0.60), substantial ($\geq$ 0.61 and $\leq$ 0.80), and almost perfect ($\geq$ 0.81 and $\leq$ 1.00). 
The Kappa value was \textit{0.571}, which means a moderate agreement level. The remaining 79\% (=11,297) of studies were applied individually.

Then there was the elimination of 13,007 studies based on the following criteria: 120 studies not written in English (EC3); 23 studies not reaching the status of a full paper (EC4); 995 not peer reviewed studies  (EC5); 7,539 studies that did not report a Secondary Study (EC6); and 4,330 studies not related to SE (EC7). At the end of this process, 1,320 secondary studies remained.

At phase \textbf{P6}, we applied the EC8 criterion to filter the studies from the top conferences and journals, there remaining \textbf{446 Secondary Studies} to be analyzed. We used this criterion to select studies published in venues with a high potential impact on academic research and industry with international coverage and eliminate studies published in predatory venues. Then, we applied the extracted data (see Section~\ref{sec:data-extraction}) for all the information needed from those studies to answer our research questions.

Finally, at phase \textbf{P7}, we conducted the data analysis and synthesis (see Section~\ref{sec:data-analysis-synthesis}) employing a qualitative and quantitative approach. 

From the total of selected studies (446), we separated and analyzed them into two samples: \textbf{126 Secondary Studies using GL} (see Appendix A) and \textbf{320 Secondary Studies that did not use GL} (see Appendix B). In Section~\ref{sec:data-analysis-synthesis}, we explain how each sample was used in our analysis.

\subsection{Data Extraction}
\label{sec:data-extraction}

We extract data relevant to answer each RQ. For RQ2, we extracted from each study the general information proposed by Da Silva et al.~\cite{daSilva:2011:IST}. For the remaining RQs, we extracted similar data to that reported by the study of Zhang et al.~\cite{Zhang:2020:ICSE}. Some data extracted from each study includes (but is not limited to): (i) names of authors, (ii) year of publication, (iii) authors' institution, (iv) institutions' country, (v) quantity of included studies, and (vi) motivations to use or reasons to avoid GL. In addition, considering each study that included GL, we extracted the following information: (i) GL type, (ii) whether the GL data is still available online.

\subsection{Data Analysis and Synthesis}
\label{sec:data-analysis-synthesis}

We employed a mixed-method approach based on both qualitative and quantitative methods to analyze data. We used a \textit{qualitative} approach when we were interested in questions about ``what'' and ``how.'' To complement this qualitative analysis, we used descriptive statistics to discuss frequency and distribution.

\subsubsection{Qualitative Approach}

\begin{figure}[h!]
\centering
\includegraphics[scale = 0.3, clip = true, trim= 100px 30px 110px 0px]{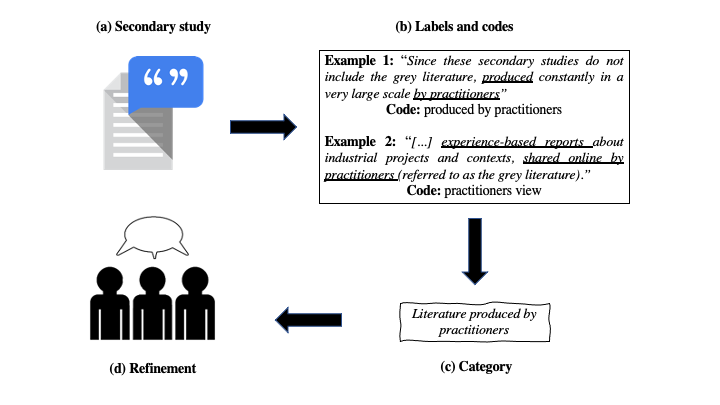}
\caption{Example of coding process used in the studies.}
\label{fig:code-categories}
\end{figure}

Our qualitative approach followed a thematic analysis technique~\cite{Cruzes:ESEM:2011}. Figure~\ref{fig:code-categories} presents a general overview of this approach. We describe it in greater details in these next points (adapted from Pinto et al.~\cite{Pinto:ICSE-SEET:2019}):

\begin{itemize}
   \item \textbf{Familiarizing ourselves with data:} Each researcher involved in the data analysis procedure read (and re-read) every Secondary Study, as expressed in Figure~\ref{fig:code-categories}-(a).
   
   \item \textbf{Initial coding:} In this step, each researcher individually added codes. We used a post-formed code. We labeled portions of text without any pre-formed code. Labels express the meaning of excerpts from the answer that represented appropriate actions or perceptions. The initial codes were temporary since they needed refinement. All identified codes were refined through a collective analysis. Figure~\ref{fig:code-categories}-(b) presents an example of this analysis. Considering this example, there are two examples of portions of text extracted in which two codes were generated, \textit{produced by practitioners} and \textit{practitioners' view}. However, these codes were classified with a single code since they have the same meaning.

   \item \textbf{From codes to categories:} Here, we already had an initial list of codes. A single researcher looked for similar codes in data. Codes with similar characteristics were grouped into broader categories. Eventually, we also had to refine the categories found, comparing and re-analyzing them in parallel. Figure~\ref{fig:code-categories}-(c) presents an example of this process. This example exhibits how the category ``definitions of Grey Literature'' emerged.
   
   \item \textbf{Categories refinement:} Having finished the previous step, we gathered a set of candidate categories. In this step (Figure~\ref{fig:code-categories}-(d)), we involved three researchers: two to evaluate all categories, and a third researcher to resolve any disagreements (if needed). Besides aiming for accurate results, we rename and regroup some categories. The third researcher, once again, was invited to review and provide comments on those categories. In the cases of any doubt, we resolved them though conflict resolution meetings.
 \end{itemize}

\subsubsection{Quantitative Approach}

We used two samples to our quantitative approach: (i) 446 Secondary Studies used to answer RQ1, RQ2 and RQ5 (\textit{precisely the reasons to avoid its use}); and (ii) 126 studies that used GL answered the questions RQ3, RQ4, RQ5 (\textit{specifically the motivations to use}), and RQ6.

We highlight that one Secondary Study could be related to more than one category found for quantitative analysis. Moreover, we calculated the percentage of answers based on the total population investigated (126 or 446). For example, to address RQ1, we considered all 446 studies and we found 150 answers which were reported in 116 studies. Thus, we calculate to this category the value of 150/446.

\section{Results}\label{sec:results}
In this section, we present the results for each of the research questions described in Section~\ref{sec:RQ}.

\subsection{RQ1. What definitions of Grey Literature are employed in Secondary Studies?}

From the 446 selected secondary studies investigated, we found 150 ones (33.6\%) presenting some GL definitions. Among them, only 34 studies used general terms such as ``grey or gray.'' On the other hand, however, 116 studies did not use any clear definition; instead, they used GL characteristics to express its definition.
We found the following categories (some studies were classified into more than one category).

\vspace{0.2cm}
\noindent
\textbf{Expressed by the types of Grey Literature (108/446 studies; 24.2\%).} Here we group studies that  expressed GL in terms of its types. For example, in the study of Do Carmo Machado et al.~\cite{SS50} pointed out that: \textit{``Gray literature herein includes \ul{technical reports} and \ul{book chapters}.''} Similarly, Irshad et al.~\cite{SS42} informed that: \textit{``New search terms were collected by browsing through Grey Literature (\ul{technical reports}, not peer reviewed articles, and \ul{webpages}).''}

\vspace{0.2cm}
\noindent
\textbf{Not peer reviewed (75/446 studies; 16.8\%).} This category groups ``non-peer-reviewed'' documents. For example, Tripathi et al.~\cite{SS34} excluded papers that did not pass through a peer review process, as quoted here: \textit{``[...] with its results, researchers and practitioners can consider both viewpoints \ul{that were non-peer-reviewed} (gray literature) [...].''} Li et al.~\cite{RQSS30} mentioned that \textit{``A publication that has \ul{not undergone a peer-review} is considered informal and not included''}, and is therefore GL.

\vspace{0.2cm}
\noindent
\textbf{Literature produced by practitioners (40/446 studies; 9\%).} This category mentioned GL as literature published in the industry but not in academic settings. For example, Garousi et al.~\cite{SS98} pointed out that: \textit{``These secondary studies do not include the grey literature \ul{produced} constantly in a very large scale \ul{by practitioners}.''} Raulamo-Jurvanen et al.~\cite{SS47} complements that: \textit{``[...] experience reports about industrial projects and contexts, \ul{shared online by practitioners} (referred to as grey literature).''}

\vspace{0.2cm}
\noindent
\textbf{Non-published literature (24/446 studies; 5.4\%).} This category refers to studies not published, such as work in progress or non-indexed works. For example, Nadal et al.~\cite{SS9} mentioned it as we quoted: \textit{``[...] five \ul{non-indexed works} considered grey literature were additionally added to the list.''} In contrast, there was an exclusion criterion for the work of Mohabbati et al.~\cite{SS70} as we pointed out: \textit{``[...] excluding `gray publications' such as short papers, \ul{works in progress, unpublished}, or non-verified literature.''}

\vspace{0.2cm}
\noindent
\textbf{Others (13/446 studies; 2.9\%).} We group here studies that employed other definitions, including the mapping study of Sharma and Spinellis~\cite{SS13} that mentioned GL as a secondary source, as quoted: \textit{``We did not limit ourselves only to the primary studies. We included \ul{secondary sources of information} and articles as and when we spotted them while studying primary studies.''} The research of Sharafi et al.
~\cite{RQSS29} attributed GL to a lack of trust: \textit{``Papers in `grey’ literature, which \ul{are not published by trusted, well-known publishers}.''}

\begin{shaded}
\noindent
\textbf{Summary for RQ1:} We found that most studies did not use the term grey/gray literature. Instead, they refer to GL in terms of its types or characteristics (literature produced by practitioners and non-published literature).
\end{shaded}

\subsection{RQ2: How is Grey Literature used in Secondary Studies?}

\subsection*{Usage of Grey Literature in Secondary Studies}
\begin{figure}
	\centering
\includegraphics[scale = 0.30, clip = true, trim= 0px 0px 0px 0px]{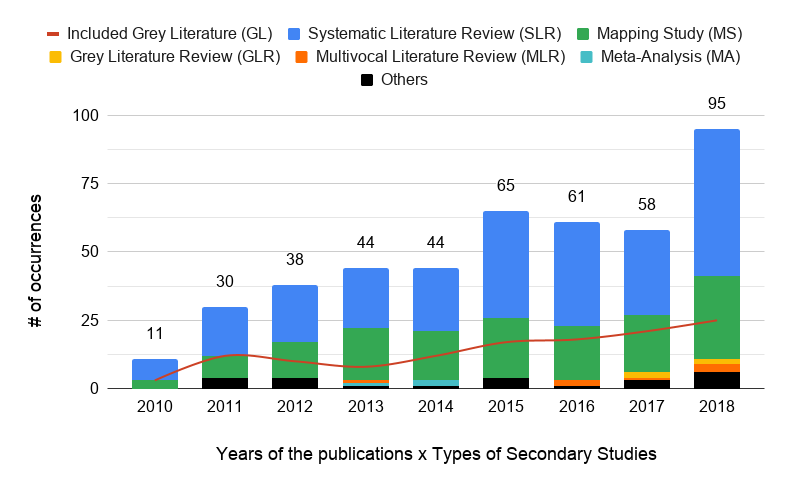}
	\caption{Distribution of each type of Secondary Study with line trend of Grey Literature use over the years.} 
	\label{fig:chart-distribution-secondarystudies-years}
\end{figure}

Figure~\ref{fig:chart-distribution-secondarystudies-years} presents the temporal distribution of 446 Secondary Studies, showing an increase in studies published over the years along with the development of trend line (represented by the red line) of studies using GL.

Overall, from the 446 selected secondary studies, we found a subset of \textbf{126 studies that used GL (28.2\%)}. To understand how our selected studies are using GL, we start by investigating the Secondary Studies methodological section, which often provides information about the search and selection procedures. As a shortcoming, we noticed that some studies did not mention such information despite the lack of studies mentioning any criteria for using (or not using) GL. Our manual analysis found GL references in the list of selected studies.

\subsection*{Grey Literature vs Types of Secondary Studies}

\begin{figure}
	\centering
	\includegraphics[width=\linewidth,clip = true, trim= 0px 20px 0px 0px]{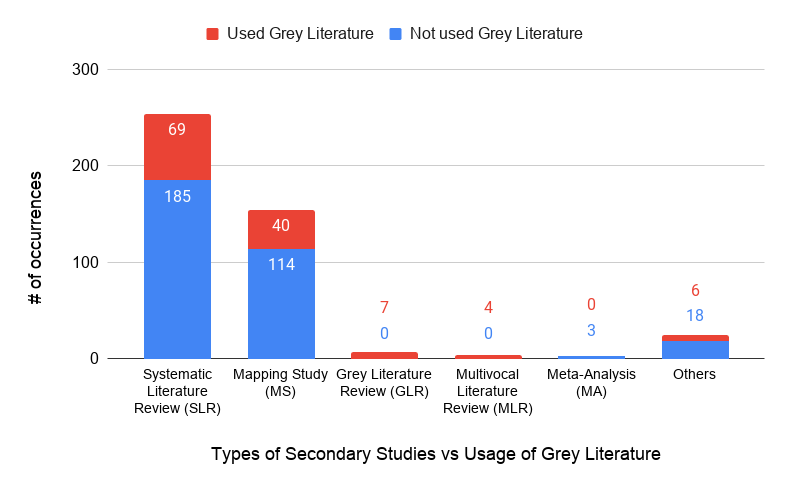}
	\caption{Distribution of the studies using and not using Grey Literature grouped by the types of Secondary Studies.}
	\label{fig:chart-gl-usage}
\end{figure}

Figure~\ref{fig:chart-gl-usage} shows a different scenario. It breaks down GL usage in terms of Secondary Study (i.e., SLR or MLR). As we could see, GL was not widely used in Secondary Studies.

Table~\ref{tab:gl-distribution-primary-studies} shows the distribution (\%) of GL between included studies according to the Secondary Study type. This table shows that for most of the Secondary Studies there was the inclusion of up to 10\% of GL studies. The studies of MLR caught our attention because, for the most of them, the GL studies represented over 50\% of the studies included.

\begin{table*}[t]
\centering
\caption{Intervals of distribution of Grey Literature included over the total of studies included by each type of Secondary Study.}
\label{tab:gl-distribution-primary-studies}
\begin{minipage}{.8\textwidth}
  \begin{tabular}{lccccc}
    \toprule
    Type of Secondary Study & <=10\% & 11-25\% & 26-50\% & 51-75\% & 76-100\%\\
    \midrule
    Systematic Literature Review (SLR) & 48 & 16 & 3 & 1 & 1\\
    Mapping Study (MS) & 31 & 4 & 4 & 1 & 0 \\
    Multivocal Literature Review (MLR) & 0 & 0 & 3 & 3 & 1 \\
    Grey Literature Review (GLR) & 0 & 0 & 0 & 0 & 0\\
    \bottomrule
  \end{tabular}
\end{minipage}
\end{table*}

\subsection*{Usage of Grey Literature to support research questions}
We also investigated to what extent GL is used to support the answers to the research questions distributed between the types of Secondary Studies. 

\begin{figure}
	\centering
	\includegraphics[width=\linewidth, clip = true, trim= 0px 0px 0px 0px]{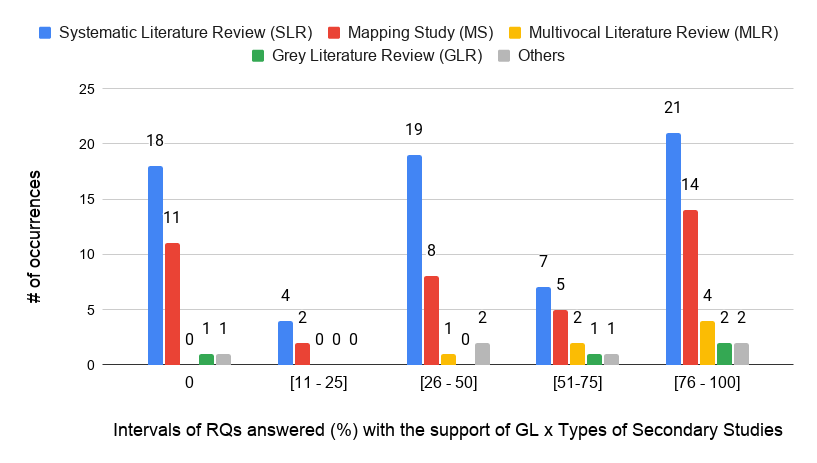}
	\caption{Coverage (\%) of RQs answered with the support of Grey Literature distributed for each type of Secondary Study.} 
	\label{fig:chart-interval-rq-answered}
\end{figure}

Figure~\ref{fig:chart-interval-rq-answered} shows the percentage of RQs answered by each type of Secondary Study using at least one example of GL. Our analysis found \textbf{95 studies (95/126 studies; 75.4\%) using GL to support answers to at least one research question}. 
However, by the individual analysis of each study, we found less than half of the studies using GL to answer more than half of the RQs. We found two interesting results: (i) we found 31 studies (24.6\%) that included GL but did not use them to answer any of the RQs, and (ii) RQs of six MLR studies (4.8\%) were not answered with the support of GL.

\begin{shaded}
\noindent
\textbf{Summary for RQ2:} We perceived that Secondary studies do not widely use GL, although there is an increase in its use over the years. We also found that 75\% of studies that cited GL, used GL evidence to answer at least one of their research questions.
\end{shaded}

\subsection{RQ3. How is Grey Literature searched, selected, and has the quality assessed in Secondary Studies?}

In this section, we investigate the procedures used to collect, select, and analyze data. Firstly, we started by investigating if the study followed any guidelines to support its research. We found that Kitchenham and Charters' guidelines~\cite{Kitchenham:2007:Guideline} were most used (67 studies) among Secondary Studies, followed by Wohlin's guidelines used to conduct snowballing~\cite{Wohlin:2014:Snowballing}. We did not find a specific guideline to conduct GLR. 

\subsection*{Search methods}

We discovered that most of Secondary Studies (62.7\%) used multiple search methods (automatic, manual, and snowballing) to search for studies. Table~\ref{tab:search-sources} describes the procedures employed with the total number of occurrences of a given procedure in column ``\#''. We found most Secondary Studies using \textit{academic search engines} (e.g., ACM Digital Library, IEEE Xplore, Scopus, Science Direct, and SpringerLink) for data collection because most of them are studies using traditional SLR and MS focused on academic findings. The use of \textit{Google Scholar} was the second most used search engine, followed by \textit{Google}. All GLRs and MLRs have used \textit{Google} as a primary search engine.

\begin{table}[H]
\caption{Procedures used to search for GL. The column \# shows the total number of studies of a given category. Note: *Automatic search.}
\label{tab:search-sources}
\begin{minipage}{.5\textwidth}
  \centering
  \begin{tabular}{lcc}
    \toprule
    Source & \# & \%\\
    \midrule
    Academic search engines* & 102 & 80.9\%\\
    Google Scholar* & 47 & 37.3\%\\
    Google* & 28 & 22.2\%\\
    Microsoft Academic Research* & 3 & 2.4\%\\
    Manual search & 31 & 24.6\% \\
    Specialized databases & 14 & 11.1\% \\
    Snowballing & 50 & 39.7\% \\
    \bottomrule
  \end{tabular}
\end{minipage}
\end{table}

Based on our sample analysis, from 10 GLR and MLR studies that intend to search for GL studies, only two of them used different sources from the traditional secondary studies to search for GL. 
These findings show that \textbf{most of the studies did not apply any particular strategy to search for GL} (only 14 studies used specialized databases, e.g., Agile Alliance, YouTube, and Stack Overflow). Zhang et al.~\cite{Zhang:2020:ICSE} also found this lack of a specific strategy to search for GL.

\subsection*{Selection criteria}
We found 20 studies did report their intention to include only peer reviewed studies. On the one hand, 13 out of 20 studies described exceptions to the inclusion criteria, most of them related to specific types of GL, e.g., technical reports (e.g.,~\cite{SS3}) and books (e.g.,~\cite{SS28}). 
In the following sections, we present the groups of inclusion and exclusion criteria that emerged from our analysis.

\begin{itemize}

\item \textbf{Inclusion.} The most common inclusion criterion was to include \textit{Specific types of GL} found in 45 studies. 
For example, Fernand\'ez et al.~\cite{SS61} include Ph.D. theses and technical reports. Tiwari and Gupta~\cite{SS5} study mentioned that: \textit{``The \ul{technical/experience reports, white papers, and books' chapters} were searched by reviewing the references of the selected papers.''}
Other studies (4 studies) used the criterion to include \textit{industrial publication}, as quoted in the work of Lewis and Lago~\cite{SS17}: \textit{``A study that is in the form of a published scientific paper or \ul{industrial publication}.''}
Other inclusion criteria were also used, but to a lesser extent, for example, Garousi et al.~\cite{SS112} specifically included GL from the \textit{Seminal source}, as mentioned here: \textit{``[...] one non-peer-reviewed technical report published by the Software Engineering Institute (SEI), which is \ul{considered a highly credible} institute for software engineering research.''} Another criterion was to consider the \textit{authors' credibility} to include specific studies, as quoted: \textit{``Publishing companies or websites \ul{suggested by experts}.''}

\item \textbf{Exclusion.} The most common exclusion criterion was to exclude \textit{Specific types of GL} found in 33 studies as the inclusion criteria. 
For example, Pedreira et al.~\cite{SS27} searched for GL. However, personal blogs or web pages were excluded from the search. Mahdavi-Hezavehi et al.~\cite{SS19} also excluded specific types of GL, as quoted: \textit{``(Exclusion criteria) The study is an \ul{editorial, position paper, abstract, keynote, opinion, tutorial summary, panel discussion, or technical report}.''}
We found other exclusion criteria but to a lesser extent, including reports with a \textit{Lack of details} (e.g., from Quora, Slideshare, or Linkedin), and \textit{Web resources without keywords from search string} (e.g., if one of the keywords was missing from the resource, it was automatically discarded). 
Another restriction was to exclude \textit{websites without text} because they considered it hard to analyze them, as pointed out in the study of Tripathi et al.~\cite{SS34}: \textit{`` If the webpage \ul{is only videos, audio, or images without text}, it should be excluded.''}
\end{itemize}

\subsection*{Quality assessment}
We discovered that \textbf{only seven studies employed specific criteria to assess the quality of the GL.} Among those studies, we found three MLRs, one GLR, and three of the other types. The GLR conducted by Soldani et al.~\cite{SS113} employed specific criteria to assess the GL combined with inclusion/exclusion criteria to filter the studies. These criteria were grouped into four groups: \textit{practical experience} measured in years of experience in the subject, \textit{industrial case} that reported previous experience on the subject, \textit{heterogeneity} of the results, and the \textit{implementation quantity} that refers to the detail in which the results were discussed.

Tom et al.~\cite{SS67} informed that, due to the diverse nature of an MLR, it was necessary to consider the particularity of each type of GL. 
They assessed the studies in terms of the position and certainty of the source, clarity, detail, consistency, and plausibility. 
In the study of Garousi et al.~\cite{SS90}, specific criteria were used to assess GL quality which covered the following aspects: authority, accuracy, coverage, objectivity, date, and significance.

\begin{shaded}
\noindent
\textbf{Summary for RQ3:} We do not found any specific guidelines to conduct a GLR, and most of the studies, even the MLR or GLR studies, used Kitchenham's or Petersen's guidelines. Moreover, few studies used a specific criteria to search for GL or used specific criteria to assess its quality.
\end{shaded}

\subsection{RQ4: What types of Grey Literature are the most frequently used in Secondary Studies?}

This section investigates the types of GL used and the producers of the GL found. We also assess the GL availability.

We perceived that most of the studies (54\%) (68/126 studies) did not classify the GL included. We used the classification of those that made it available. Since they were not classified, we had to classify them according to our interpretation, for instance, by reading the reference meta-information or accessing the link when it is available. However, there were some cases where we could not perform this assessment process (e.g., on a website/link no longer available, discontinued blog post, vague references). We classified these examples of GL as ``Unknown.''

The Secondary Studies mentioned a total of 1,314 examples of GL included. However, from this amount, when investigating the list of references of those studies, we retrieved only 1,273 GL studies (41 were missing). Moreover, from this list, we removed 25 peer reviewed studies erroneously classified as GL. At the final, 1,246 GL studies remaining, distributed into 21 types.

\begin{figure*}[h]
    \centering
    \includegraphics[scale= .55]{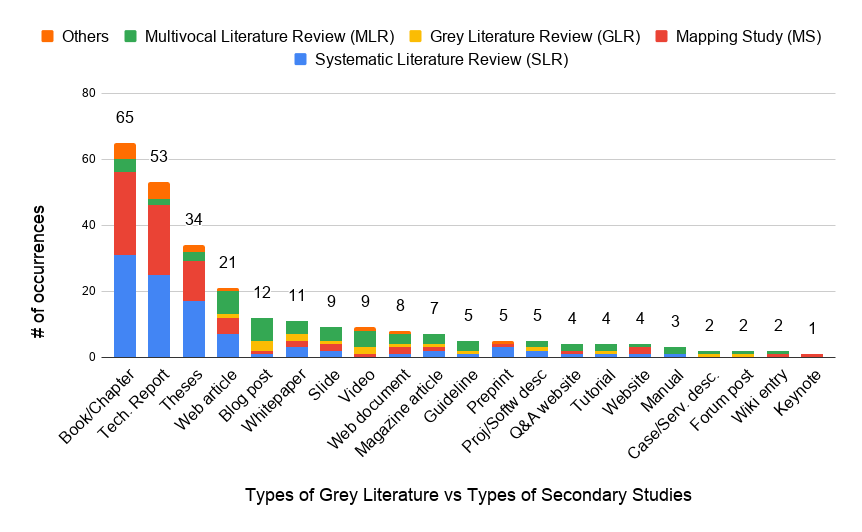}
    \caption{Distribution of the types of Grey Literature used between the Secondary Studies. Here it is evident, that 65 studies used books/chapters; 53 studies used technical reports.}
    \label{fig:occurences-types-gl}
\end{figure*}

\subsection*{Types of Grey Literature in Secondary Studies}

Firstly, Figure~\ref{fig:occurences-types-gl} shows the GL types most commonly used distributed between the types of Secondary Studies. The figure shows that \textit{books/chapters} 
were the most common type of GL found among the Secondary Studies, followed by \textit{technical reports}, 
\textit{theses}, 
\textit{web articles}, 
\textit{blog posts}, 
and \textit{whitepapers}. 

Secondly, we analyzed all the 21 types of GL found in the Secondary Studies and classified them according to the ``shades'' of grey proposed by Garousi et al.~\cite{Garousi:2019:IST} (see Figure~\ref{fig:shades-gl}). 
We found four GL types on the first tier, nine on the second tier, and eight on the third tier, showing that most GL types used have a medium level of control and expertise.

Blog posts (third tier) were most commonly found in the MLR and GLR studies. We also found many web/news articles, whitepapers, and descriptions of projects, software, and tools (second tier).

\begin{figure*}[hp]
    \centering
    \captionsetup{justification=centering}
    \includegraphics[scale= .45, clip = true]{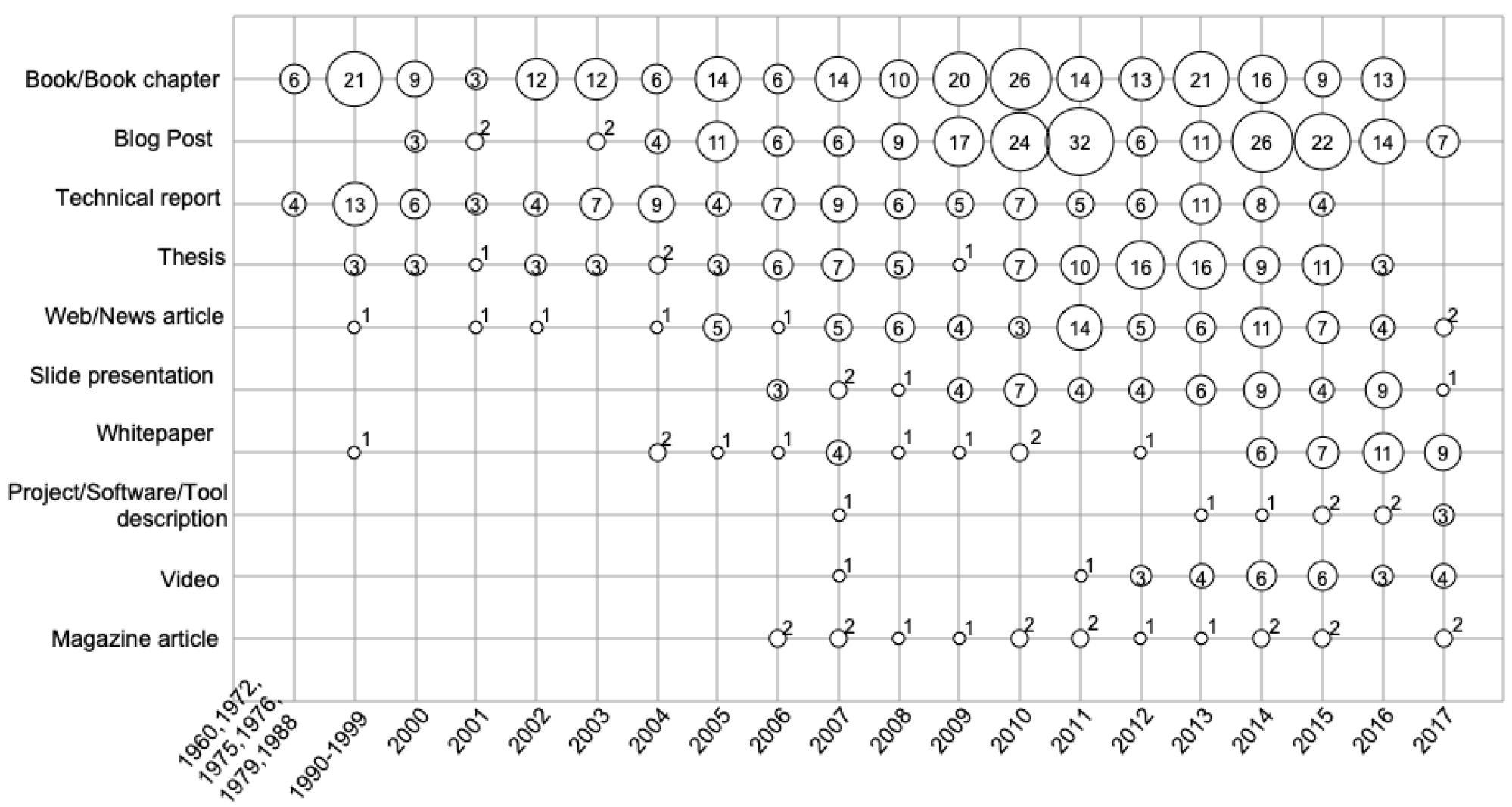}
    \caption{Distribution of each type of Grey Literature found in the Secondary Study distributed over the years of publication.}
    \label{fig:occurences-types-gl-years}
\end{figure*}

Figure~\ref{fig:occurences-types-gl-years} shows the evolution of GL types usage over the years. Due to constraints related to space, we selected the ten most used types of GL. 
Due to the small amount of GL arising in these years, in the first column, we combined the types of GL published between 1960 and 1988. Further still, in the second column, we combined the GL published from 1990 to 1999. 
Overall, we found a growth in GL's use since 2009, which was mainly driven by blog posts, theses, and web/news articles. The use of books/chapters did not change significantly over the years. On the other hand, Secondary Studies are lately adopting whitepapers, videos, and descriptions of projects, software, and tools. In particular, blog posts start to be frequently published from 2000, and their inclusion as GL steadily increased over the years till the point it became one of the most used GL documents.

\subsection*{Grey Literature Producers}
One of the criteria proposed by Garousi et al.~\cite{Garousi:2019:IST} to conduct the quality assessment in GL is to check the reputation of the authors and/or publishing organization. However, the information on who produces each example of GL has not always available. In these cases, we used our interpretation to fill this gap (e.g., accessing the links). To classify different producers of GL, we followed Maro et al. as a reference~\cite{MARO:JSS:2018}.

We analyzed all the 1,246 GL to identify: (i) who the producer was and (ii) which GL types each producer was related to. Our first analysis is related to the data present in Table~\ref{tab:producers}, which shows the total number of occurrences of contents produced by each type of GL producer in the column ``\#''. Three types of producers (\textit{Consultants / Companies, Academia,} and \textit{Practitioners}) caught our attention, responsible for producing almost 80\% of the GL included. Other important information is that for 13.2\% of the GL studies it was not possible to determine its producer.

\begin{table}[H]
\caption{Grey Literature Producers. The column \# shows the total number of Grey Literature of a given type of producer.}
\label{tab:producers}
\begin{minipage}{.5\textwidth}
  \centering
  \begin{tabular}{lcc}
    \toprule
    Producer & \# & \%\\
    \midrule
    Consultant/Company & 391 & 31\%\\
    Academia & 361 & 28.6\%\\
    Practitioner & 230 & 18.2\%\\
    Tool vendor & 67 & 5.3\%\\
    Standardisation Body & 10 & 0.8\%\\
    Agency & 7 & 0.6\%\\
    * Others & 14 & 1.1\%\\
    * Unknown & 166 & 13.2\%\\
    \bottomrule
  \end{tabular}
\end{minipage}
\end{table}

\begin{figure*}[!th]
\centering
\captionsetup{justification=centering}
\includegraphics[scale = .55]{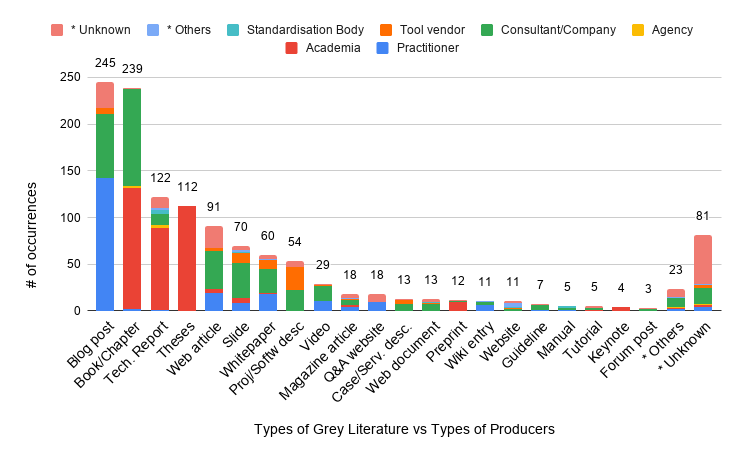}
\caption{Distribution of each type of Grey Literature between the diverse types of Producers.}
\label{fig:types-gl-vs-producers}
\end{figure*}

Our second analysis is related to Figure~\ref{fig:types-gl-vs-producers}, which shows the relationship between the GL types and producers types. We noted that producers from \textit{Consultants} / \textit{Companies} were the ones that most produced GL content, as one could expect, related to diverse GL types, but the most common were \textit{Books / chapters}, \textit{Blog post}s, and \textit{Web articles}. Moreover, what caught our attention is that \textit{Consultants / Companies} were the ones that produced the most content of \textit{Web articles}, \textit{Slides}, and \textit{Videos}. The second was the \textit{Academia}, mainly related to \textit{Theses}, \textit{Technical reports}, and \textit{Books / chapters}. The third was the \textit{Practitioners}, where the production was most related to \textit{Blog posts}, \textit{Web articles}, and \textit{White papers}.

\subsection*{Availability of the Grey Literature}

Kitchenham et al.~\cite{Kitchenham:2015:ESE:BOOK} argued on the importance of traceability in a Secondary Study so that the data would be available to others in the future. For this reason, we investigated all the 1,246 GL references included in the Secondary Studies (n=126) to verify: (i) If the URL of each GL was informed; (ii) If each URL reported is still working, that means we did not have any problem accessing it, and it was directed to the correct link of the source; and (iii) If were reported any access problems with the URLs.

Our findings show that 24.8\% of GL sources (309/1,246) only presented the reference title but \textit{did not report a URL}. From the remaining that \textit{URLs were informed}, we found 23.7\% (295/1,246) that exhibited \textit{access problems} (e.g., server not found, page not found), leaving 51.5\% (642/1,246) of GL references \textit{still available}.

We also investigated the producer of each GL source, showing that among the sources in which URLs were not informed, more than half was produced by \textit{Academia}, and among the URLs that returned some access problem, more than 30\% of the items were from \textit{Consultant / Company}.

\begin{shaded}
\noindent
\textbf{Summary for RQ4:} The most commonly found GL types among the studies were books/chapters, technical reports, and theses, produced mainly by academia and consultants/companies. The practitioners produced most of the blog posts. Our investigation for GL references showed that, unfortunately, a quarter of the GL URLs were not reported, a little more than half of the GL URLs are still working, and for 23.7\% of GL, we found some impediments to access.
\end{shaded}

\subsection{RQ5: What motivates researchers to use/avoid Grey Literature?}

In this question, we used two samples for our analysis. For the motivations \textit{to use} GL, we investigated only the 126 selected studies; we investigated all of the 446 Secondary Studies for the reasons to avoid. We maintained the studies that did not use GL in the analysis because some reported the reasons to avoid GL. Some studies, though, did not explain. We did not consider them in this research question.

\subsubsection*{Motivations to use Grey Literature}
\label{sec:motivations_use_gl}

We noted that the discussion in favor of using GL is often provided in the search process or in the Secondary Study's inclusion criteria. For the GLR and MLR studies, it was common to find some of this information in the introduction as well. 
We found 35 out of 126 Secondary Studies that presented motivations to use GL. After grouping the answers, we presented the categories found in Table~\ref{tab:motivations-use-GL} and then discuss them.

\begin{table}[H]
\caption{Motivations to use Grey Literature. The column \# shows the total number of studies of a given category.}
\label{tab:motivations-use-GL}
\begin{minipage}{.4\textwidth}
      \centering
  \begin{tabular}{lcc}
    \toprule
    Motivation & \# & \%\\
    \midrule
    To identify more studies & 16 & 12.7\%\\
    To incorporate practitioners' point of view & 10 & 7.9\%\\
    To reduce publication bias & 5 & 4\%\\
    Others & 4 & 3.2\%\\
    No motivation was given & 91 & 72.2\%\\
    \bottomrule
  \end{tabular}
\end{minipage}
\end{table}

\vspace{0.2cm}
\noindent
\textbf{To identify more studies (16/126 studies; 12.7\%).} This motivation is the most common among the Secondary Studies. T\"uz\"un et al.~\cite{SS26} state that they: \textit{``searched for company journals, gray literature, conference proceedings, and the internet, which \ul{led us to new papers} that we could not identify in our regular search.''} The study of Carrizo et al.~\cite{SS52} reviewed books and Ph.D. theses to identify more studies. Besides, Garousi and M\"antyl\"a~\cite{SS126} investigated the ideal moment to automate, and what to automate in software testing and decided to include GL because \textit{``the \ul{academic} studies on the \ul{topic were rare}.''}

\vspace{0.2cm}
\noindent
\textbf{To incorporate the practitioners' point of view (10/126 studies; 7.9\%).}
This category was the second most common motivation among the studies. Tripathi et al.~\cite{SS34} state that: \textit{``we \ul{need to incorporate the practitioners' point of view}, which is shared through internet channels in the form of webpages and whitepapers.''} Other studies~\cite{SS47,SS112,SS126} included GL because they wanted to hear the practitioners' ``voice''.

\vspace{0.2cm}
\noindent
\textbf{To reduce publication bias (5/126 studies; 4\%).} 
Some studies were motivated to include GL to reduce publication bias. As mentioned by Patel and Hierons~\cite{SS6}, \textit{``including grey literature is an important step \ul{to combatting publication bias}.''} The study of Rizvi et al.~\cite{SS8} mentioned that \textit{``We should point out that grey literature, such as the organization of white papers and lessons learned, were reviewed manually \ul{to address bias in paper selection}.''}

\vspace{0.2cm}
\noindent
\textbf{Others (4/126 studies; 3.2\%).} We group here the studies that employed other motivations. For instance, Soldani et al.~\cite{SS113} pointed that companies are working day-by-day on the design, development, and operation of research interest, as also witnessed by the high number of GL items on the topic. Another motivation was the use of a \textit{trustworthy source} such as Chen and Babar's study~\cite{SS109}, which decided to manually search the technical reports from the Software Engineering Institute (SEI) because they considered it a trustworthy source, as pointed out here: \textit{``\ul{SEI's} series of technical reports is the \ul{main channel of grey literature in the research area}.''}

\subsubsection*{Reasons to avoid Grey Literature}
\label{sec:motivations_avoid_gl}

We also analyzed the reasons for the studies to \textit{avoid using} GL. Distinct from the motivation to use it, in this question, we also looked at the 446 studies, both the studies that \textit{used} and did \textit{not use} GL. It was most common to find information in the studies' inclusion/exclusion criteria section and the section on the threats to validity. We found 28 out of 446 Secondary Studies that presented reasons to avoid GL. After grouping the answers, we present the categories found in Table~\ref{tab:reasons-avoid-GL} and then discuss them.

\begin{table}[H]
\caption{Reasons to avoid Grey Literature. The column \# shows the total number of studies of a given category.}
\label{tab:reasons-avoid-GL}
\begin{minipage}{.5\textwidth}
  \centering
  \begin{tabular}{lcc}
    \toprule
    Reason & \# & \%\\
    \midrule
    Lack of Quality & 23 & 5.1\%\\
    Hard to identify Grey Literature & 3 & 0.7\%\\
    Others & 2 & 0.4\%\\
    No reason was given & 418 & 93.7\%\\
    \bottomrule
  \end{tabular}
\end{minipage}
\end{table}

\vspace{0.2cm}
\noindent
\textbf{Lack of Quality (23/446 studies; 5.1\%).} The most common motivation to avoid GL concerns the lack of quality that may affect the validity of the results. This category consists of a set of subcategories related to the nature of lack of peer review process, some validity constraints, and the reliability of this type of source. The lack of peer review processes was the most common subcategory found to avoid GL as Alkhanak et al.~\cite{RQSS02} pointed out to explain the motivation to exclude GL: \textit{``[...] there could be a \ul{threat} associated with Grey literature, that \ul{no peer-reviewed process} might have been adopted''}. Another motivation that GL might negatively affect the validity because it is not externally valid as pointed out in Vilela et al.~\cite{RQSS22}: \textit{``[...] the external validity depends on the identified literature: if the identified literature is \ul{not externally valid}, neither is the synthesis of its content [...].''}

\vspace{0.2cm}
\noindent
\textbf{Hard to identify Grey Literature (3/446 studies; 0.7\%).} Some authors eliminated GL because they considered it hard to identify GL in their research interest (e.g., ~\cite{RQSS80}). Moreover, Lane and Richardson~\cite{RQSS14} excluded GL \textit{\ul{to make the SLR more straight-forward} and \ul{repeatable} but at the cost of potentially excluding valuable studies}.

\vspace{0.2cm}
\noindent
\textbf{Others (2/446 studies; 0.4\%).} We group here the studies that employed other motivations. For instance, Arvanitou et al.~\cite{RQSS11} eliminated GL because it considered that they had included the majority of ``good quality'' studies published in the selected venues, and increasing the number of studies would seriously threaten the feasibility of their investigation. A study by Alkhanak et al.~\cite{RQSS02} mentioned two reasons to exclude GL: lack of technical details and the lack of quality.

\begin{shaded}
\noindent
\textbf{Summary for RQ5:} According to our findings, it was possible to categorize the studies' motivation to use and reasons to avoid GL. On the one hand, studies were motivated to use GL to identify more studies, incorporate practitioners’ points of view, and reduce their studies' publication bias. On the other hand, they avoided using GL because of lack of quality (e.g., not peer-review and validity constraints).
\end{shaded}

\subsection{RQ6: How do researchers perceive the use of Grey Literature?}

In this question, our sample analysis of 126 Secondary Studies both the \textit{benefits} and \textit{challenges}, focusing only on the studies using GL. Not all Secondary Studies have data that could be used to answer this research question.

\subsubsection*{Benefits perceived}\label{sec:benefits_gl}

We analyzed the benefits mentioned in the studies (n=126) concerning the use of GL. We found only 13 Secondary Studies that present answers to this question. In the following, we present categories of the benefits found.

\vspace{0.2cm}
\noindent
\textbf{Provision of practical Evidence (13/126 studies; 10.3\%).}
This category indicates that GL brings practical experience. Usually, SLRs are based on academic papers. However, in some areas, this information is not enough to assess a topic of interest. For example, Raulamo-Jurvanen et al.~\cite{SS47} highlight: \textit{``[...] the importance of grey literature for topics where the \ul{``voice of practice''} is broad (and more active than academic literature).''} In the same way, Soldani et al.~\cite{SS113} point out: \textit{``[...] grey literature studies can be valuable to shed light on yet uncharted areas of software engineering research, especially when such areas are seeing \ul{massive industrial adoption}.''}

\vspace{0.2cm}
\noindent
\textbf{Knowledge acquisition (9/126 studies; 7.1\%).} According to this category, the white literature is not enough to cover some topics completely, missing important knowledge on specific research areas. For this reason, some studies stressed the importance of GL to fill this gap. For instance, Garousi et al.~\cite{SS112} informed us that: \textit{``[...] if we were to exclude the grey sources from the pool, we would simply \ul{miss a major pile of experience and knowledge} from practicing test engineers on the topic.''} Another study by Garousi et al.~\cite{SS98} compares MLRs with SLRs and says: \textit{``We believe that conducting an MLR in the area of TMA/TPI will be more useful compared to an SLR since there is a \ul{large body of knowledge}.''}

\vspace{0.2cm}
\noindent
\textbf{Makes academic studies more interesting (6/126 studies; 4.8\%).}
Making academic work more interesting for practitioners is what reveals this category. For example, Garousi and M\"antyl\"a's study~\cite{SS126} reports: \textit{``[...] grey literature in SLR studies is insightful, and thus the authors recommend including it when the topic has a \ul{low number of academic studies but high practitioner interest}.''} Also, Raulamo-Jurvanen et al.'s study~\cite{SS47} adds: \textit{``Grey literature seems to have its place in SE, not only in serving the practitioners but also in \ul{providing an interesting aspect into academic studies}.''}

\vspace{0.2cm}
\noindent
\textbf{Coverage of different results from scientific studies (3/126 studies; 2.4\%).} This category indicates that GL brings results uncovered by scientific literature to the discussion, as the study of Tripathi et al.~\cite{SS34} points out: \textit{``Since the research on software startups, especially in the area of RE, is \ul{still in the nascent stage}, we need to incorporate the practitioners’ point of view, which is shared through the internet channel in the form of webpages and whitepapers.''} Also, Soldani et al.~\cite{SS113} complements: \textit{``with grey literature studies can be valuable \ul{to shed light on yet uncharted areas} of software engineering research, especially when such areas are seeing massive industrial adoption.''}

\vspace{0.2cm}
\noindent
\textbf{Easy to access and read (1/126 studies; 0.8\%).} According to this category, GL provides for ease of access, as Raulamo-Jurvanen et al.~\cite{SS47} point out: \textit{``[...] grey literature is \ul{freely and easily} available for the public.''}

\subsubsection*{Challenges perceived}\label{sec:challenges_gl}

We analyzed the challenges mentioned in the studies into the use of GL (n=126). We found only 14 Secondary Studies that present answers to this question. In the following, we present the categories of challenges found.

\vspace{0.2cm}
\noindent
\textbf{Non-structured information (8/126 studies; 6.3\%).} This category brings evidence about how GL can pose great difficulties for assessment due to its lack of format and undisciplined characteristics in its structure and writing. For example, the study of Williams~\cite{SS122} reports: \textit{``blog articles and much grey literature differ in that \ul{they are varied in their structure and the formality of their language}.''} Soldani et al.~\cite{SS113} complements the position: \textit{``This is mainly because grey literature \ul{lacks a unique format acknowledged} across all sources and available data.''}

\vspace{0.2cm}
\noindent
\textbf{Epistemological problem (5/126 studies; 4\%).} According to this category, an issue in GL is experts' opinions without the support of empirical evidence about the subject. This discloses an experience of personal opinion without providing a reliable source. It was reported in the work of Garousi and M\"antyl\"a~\cite{SS126}: \textit{``it may be that some practitioners could simply be repeating the ideas/opinions that they had heard from other practitioners. Thus, as the source of knowledge was not typically revealed in GL, we are faced with an \ul{epistemological problem}. We do not know how we know what we know.''} In the same way, Garousi et al.~\cite{SS112} pointed out: \textit{``We found that sources of evidence in grey literature were often \ul{opinion or experience based} rather than relying on systematic data collection and analysis as done in scientific papers.''}

\vspace{0.2cm}
\noindent
\textbf{Time-consuming (4/126 studies; 3.2\%).} This category indicates that the amount of literature included by GL sources demands an effort not foreseen in a common SLR. Raulamo-Jurvanen et al.~\cite{SS47} pointed out: \textit{``Screening of grey literature sources can be a \ul{time-consuming} process since usually there is no applicable abstract or summary available.''} Also, Garousi and K\"u\c{c}\"uk's~\cite{SS90} state: \textit{``Since there is a vast grey literature as well as a large body of research studies in this domain, \ul{it is not practical} for practitioners and researchers to locate and synthesize such a large literature.''}

\vspace{0.2cm}
\noindent
\textbf{Difficulty in measuring quality (2/126 studies; 1.6\%).} This category arises from the difficulty of assessing GL quality. For example, Garousi and K\"u\c{c}\"uk~\cite{SS90} pointed out: \textit{``[...] we discovered that it is very difficult to uniquely \ul{measure the quality} of grey literature when conducting a systematic, controllable and replicable secondary study.''} Garousi and M\"antyl\"a~\cite{SS126} complemented: \textit{``This suggests that requirements placed on \ul{formal publishing actually increase the amount of empirical evidence} in software engineering.''}

\vspace{0.2cm}
\noindent
\textbf{Others (2/126 studies; 1.6\%).} Here we group studies that adopted other challenges. According to Anjum and Budgen~\cite{RQSS92}: \textit{``The verification of references proved a troublesome process as more references were taken from the grey literature, often being provided on web sites. As a consequence, many were either \ul{not available} or had changed their URLs.''} This is an issue also seen during our study.
Turner et al.~\cite{SS124} mentioned the difficulty in searching or finding information on GL, as pointed out here: \textit{``The identification of ‘grey literature’, however, may be more problematic due to \ul{the digital libraries and search engines used} and the lack of available benchmarks to use for validation.''}

\begin{shaded}
\noindent
\textbf{Summary for RQ6:} Our investigation found several benefits of GL uses. The most common was for providing practical evidence, followed by knowledge acquisition. The studies did not widely mention challenges, but the most common was the lack of structured information. This makes it difficult to retrieve information from a GL source. Another challenge was an epistemological problem concerning the lack of GL reliability.
\end{shaded}

\section{Discussion}\label{sec:discussion}

In this section, we revisit our main findings, discussing some of them, and relating them to our closest related works.

\vspace{0.2cm}
\noindent
\textbf{RQ1 overview.} 
As the investigations of GL in SE research are recent, we believe that our findings of GL definitions are compatible with an area still under development and that needs more in-depth studies. For instance, most investigated studies did not explicitly use the term grey or gray, making it difficult to find GL's common definition. This finding was also identified by the tertiary study of Zhang et al.~\cite{Zhang:2020:ICSE}. The same situation was found in Sch\"{o}pfel and Prost's study~\cite{Schopfel:2020}, leaving the reader to guess what GL is and what it is not. We highlighted that Sch\"{o}pfel and Prost's~\cite{Schopfel:2020} findings supported some of our definitions. For instance, ``unpublished works'' and ``non-peer-reviewed studies.''

We agree with Zhang et al.~\cite{Zhang:2020:ICSE} that misunderstandings may influence the use of GL in SE without a common definition, as we discussed in RQ4. Recently, Garousi et al.~\cite{Garousi:2020:book} proposed a definition of GL in SE research that states \textit{``Grey Literature in SE can be defined as any material about SE that is not formally peer reviewed nor formally published.''} We believe that future works will benefit from this definition.

\vspace{0.2cm}
\noindent
\textbf{RQ2 overview.} 
\textit{(i)} Our investigation showed an increase in Secondary Studies published over the years, together with the increase in studies using GL. Even though GL has not been used extensively in Secondary Studies (126/446; 28.2\%), it stands out as an important source of evidence, as shown in our findings for RQ5 and RQ6, and in previous studies~\cite{Garousi:2016:EASE,Neto:2019:ESEM,Zhang:2020:ICSE}, that showed a diversity of benefits by using GL.

\textit{(ii)} We also investigated 1,246 GL data identified among the 126 Secondary Studies included, representing <21\% of all the 446 included studies. Nevertheless, a considerable amount taking into consideration that GL use is recent in SE research. Zhang et al.~\cite{Zhang:2020:ICSE} also found a similar ratio (22\%), while Yasin et al.~\cite{Yasin:Thesis:2020} found a different picture: 76\% of the Secondary Studies included GL. This difference occurred because the study of Yasin et al.~\cite{Yasin:Thesis:2020} considered conference proceedings and workshop papers as types of GL. In this regard, Garousi et al.~\cite{Garousi:2020:book} pointed out that conferences accept all submitted papers with no peer review in some disciplines. However, highly ranked SE conferences usually have established peer review processes. Thus, conference proceedings and workshop papers do not, in general, get treated as GL in SE research.

\textit{(iii)} Moreover, from the 126 studies, 75.4\% of them used GL to support an answer to their research questions, showing the importance of GL evidence to contributing with the findings of Secondary Studies. In this regard, our findings did not corroborate with previous studies. For instance, Zhang et al.~\cite{Zhang:2020:ICSE} found that only 25\% of studies used GL to evaluate their conclusions, while Yasin et al.~\cite{Yasin:Thesis:2020} mentioned that only 9.2\% of the GL was used to support the findings. We found a different interpretation of GL \textit{usage} between the three studies. While our study interprets this usage by analyzing each research question, Zhang et al. wanted to determine if GL was used to evaluate the conclusions.

\vspace{0.2cm}
\noindent
\textbf{RQ3 overview.}
\textit{(i)} We observed that the majority of studies had used academic search engines to perform their search. What caught our attention was that of all the GLR and MLR studies, which are naturally inclined to seek GL, used Google's general search engine and, and only 10\% of them used a more specific source (e.g., YouTube, Stack Overflow, Blogs, and Twitter) or used different forms to search for GL. For the last case, two studies specifically caught our attention. The first one, conducted by Soldani et al.~\cite{SS113}, specifically searched on GL using Google, Bing, DuckDuckGo, Yahoo!, and Webopedia. The second one, the work conducted by Williams~\cite{SS122}, used reasoning makers to search for rigorous blog articles. This finding agrees with Zhang et al.~\cite{Zhang:2020:ICSE}, who found that researchers are not adopting specific strategies to search for GL. We emphasize that future research needs to focus on using relevant and specific GL sources to the investigated topic, avoiding retrieving a large amount of sometimes irrelevant data using Google. It was perceived as a challenge to the study~\cite{SS47}, that pointed out: \textit{``Screening of grey literature sources can be a time-consuming process.''}

\textit{(ii)} Some investigated studies used inclusion criteria to specific GL types, even they have mentioned to include only peer-reviewed studies. This conflict could result from the difference in interpretation of the GL types, as discussed in the RQ4 overview. Another criterion was to included seminal sources, for instance, those provided by SEI. Concern the exclusion criteria, most of them were also related to GL types, for example, blogs, personal web pages, and videos.

\textit{(iii)} Interestingly, only seven studies (5.5\%) employed criteria to assess GL. Al-Baik and Miller~\cite{SS58} mentioned that there is a lack of guidelines to assess GL. Unfortunately, these seven studies did not employ the existing quality assessment criteria proposed by Garousi et al.~\cite{Garousi:2019:IST}, which is currently the state-of-the-art method for assessing GL.

We consider the lack of quality assessment approaches for GL as a problem because the nature of GL is different from peer-reviewed studies. Loading criteria for only one side might compromise the evaluation of the other. However, another problem was raised: it is not easy to use a single type of assessment because of the different forms of GL, as pointed out in the study of Tom et al.~\cite{SS67}.

\vspace{0.2cm}
\noindent
\textbf{RQ4 overview.}
\textit{(i)} Our research found 21 types of GL used in Secondary Studies. The most common among the studies were books/chapters, technical reports, and theses. Zhang et al.~\cite{Zhang:2020:ICSE} found similar characteristics in terms of GL's most common types, but our study differs in terms of the proportions found. For example, Zhang et al. mentioned that technical reports are present in almost 66\% of the studies (we found 42\%), blog posts in 22\% (we found 9\%), books/chapters in 22\% (we found 54\%), and theses in 17\% (we found 26\%). 

In our investigation, we found difficulty in interpreting GL types. For example, studies of Irshad et al.~\cite{SS42} and Turner et al.~\cite{SS124} considered technical reports as GL, different from Tahir et al.~\cite{SS21}. Therefore, we agree with Bonato~\cite{Bonato:SearchingGLBook:2019} that a lack of GL definition is the cause of the difficulty in its interpretation.

\textit{(ii)} In our analysis of GL producers, we perceived an increase in the importance of GL produced by \textit{Consultants} / \textit{Companies} and \textit{Practitioners} over the years. They represent, respectively, 31\% and 18.2\% of the content found. \textit{Academia} had 28.6\%. Our findings did not corroborate with Yasin et al.~\cite{Yasin:Thesis:2020}, who placed Academia first (38.3\%). As we pointed in the RQ2 Overview, this difference occurred because Yasin et al. considered conference proceedings and workshop papers as GL.

\textit{(iii)} We found many URLs used to reference GL were no longer available, which reduces GL's value to other researchers and the credibility of the study that cited it.

\vspace{0.2cm}
\noindent
\textbf{RQ5 overview.} \textit{(i)} Approximately 30\% of the studies clearly state their motivations to use GL. We organized these studies into four categories. We identified that these categories were similar to the categories found by Zhang et al.~\cite{Zhang:2020:ICSE}, namely the following: to seek more related research, to understand the views of the practitioner’s community, and to avoid publication bias. However, differently from Zhang et al., we identified that only 9.7\% of the 446 studies described the motivation to avoid GL. We organized these studies into three categories.  

Two motivations to use GL caught our attention: (i) ``To incorporate the practitioners' point of view'' because, for many years, researchers have been calling for the importance of incorporating industry evidence for SE research (e.g., in~\cite{Kitchenham:2004:ICSE}). However, only 8.7\% of studies mentioned this motivation; and (ii), the category ``To reduce publication bias'' has been discussed through several areas of knowledge (e.g., in SE~\cite{Garousi:2019:IST,Zhang:2020:ICSE}, Medicine~\cite{Paez:2017:Medicine}, and Nutrition~\cite{Adams:2016:Health}). It was found to some degree in the secondary studies investigated.

\textit{(ii)} We found few studies presenting their reasons for avoiding GL use. The main cited reason was the ``Lack of Quality'' of the studies, usually related to the lack of formal peer review processes for publication. This reason was also found in survey research with Brazilian SE researchers~\cite{Kamei:SBES:2020}. 

We also found that SE researchers investigating a well-established research field tend to avoid the use of GL because of the availability of a large number of peer-reviewed papers, as pointed out in the study of Vallon et al.~\cite{RQSS110}. On the other hand, the lack of studies on a new research topic motivates GL use. For instance, Garousi and K\"u\c{c}\"uk~\cite{SS90} noted that academic research on the topic of microservices was still at an early stage. However, companies were working daily on the design, development, and operation of the field, resulting in a considerable GL on the topic.

Moreover, we identified two trade-offs between the motivations to use and the reasons to avoid GL: (i) some researchers claimed that the motivation to use GL is the possibility to include practitioner experience, as seen in~\cite{SS9,SS4}. However, others tend to avoid its use because they were worried about the study reliability~\cite{RQSS19,RQSS22}; and (ii), there was the motivation to use GL to reduce publication bias~\cite{SS6}. However, again, concerns related to GL studies quality did some studies to question its credibility. To deal with those trade-offs, we recommended the set of criteria to assess the GL credibility found in our previous study~\cite{Kamei:SBES:2020} by selecting GL sources retrieved from renowned producers or cited by others.

\vspace{0.2cm}
\noindent
\textbf{RQ6 overview.} \textit{(i)} Only a few studies (15\%) reported benefits in the use of GL. We found five categories in which the studies were placed. Comparing our findings of the benefits with previous studies, we have: (i) the tertiary study conducted by Zhang et al.~\cite{Zhang:2020:ICSE} found four categories similar to ours, namely: to seek more relevant research, to avoid publication bias, to understand the views of the practitioner's community, and to explore unchartered research areas. Zhang et al.~\cite{Zhang:2020:ICSE} pointed out one more category not identified in our work: to compare different perspectives between researchers and practitioners; and (ii) the review presented by Rainer and Williams~\cite{Rainer:2019:JSEP} also corroborate with all of our five categories.

\textit{(ii)} Also investigating the challenges in GL use, we identified five categories. As for benefits, only a few studies (11.1\%) made clear the challenges of using GL. Comparing our findings with previous studies, we have: (i) four categories, out of the five, have similar categories to Zhang et al.~\cite{Zhang:2020:ICSE}, namely: noise in GL, paucity in ways of obtaining reliable GL, difficulty in quality assessment, and uncertain availability of GL. Zhang et al.~\cite{Zhang:2020:ICSE} present one more category: Differences in the understanding of GL definition; and (ii) all of our identified categories have similar categories to Rainer and Williams~\cite{Rainer:2019:JSEP}. These authors perceived one more category: the lack of a mechanism to control the contents' variability. 

We perceived some findings of the benefits and challenges to be contradictory. They are part of the trade-off between white literature and GL. For instance, on the one hand, GL provides knowledge acquisition and practical evidence. On the other hand, the epistemological problem related to lack of reliability arises. In part, these trade-offs were expected, but they also show the need for further investigation on improving the use of the content provided and to better deal with it. For this reason, we proposed in the next section some recommendations to deal with this problem.

\section{Challenges for dealing with Grey Literature}\label{sec:challenges-recommendations}

This section presents some challenges identified based on the Secondary Studies investigated and our experience-based with this research. First, we describe the challenge. In the following, we present potential ways to address or some existent proposals on how to deal with them, as we describe here:

\vspace{0.2cm}
\noindent
\textbf{Challenge 1: Lack of Grey Literature definition and misunderstanding about its types}. Our investigation for RQ1 found little agreement exists about GL definition, corroborating with the study of Zhang et al.~\cite{Zhang:2020:ICSE}. Instead, we observed that most of the studies did not explicitly mention ``GL''. In 2020, Garousi et al.~\cite{Garousi:2020:book} proposed a definition for GL in SE. Thus, as the formal concept of GL is recent, it is not yet widespread. We suggest that this lack of agreement on the unique definition for GL introduces a bias. Accordingly, different sources can be interpreted differently to be or not classified as a GL type. For instance, while Neto et al.~\cite{SS54} considered Ph.D. and master theses as a peer-reviewed source, the study of Rodr\'iguez-P\'erez et al.~\cite{SS83} did not consider this to be so. The same conflict was found in interpreting books and book chapters as GL types (e.g.,~\cite{SS56}) or not (e.g.,~\cite{SS97}).

\MyBox{\noindent \textit{Potential way(s) to address:} We considered it essential to clarify what GL is about and the types of GL included (or excluded) to make clear decisions employed, avoiding using only its characteristics. We also recommend using Garousi's definitions~\cite{Garousi:2020:book} that stated, \textit{``Grey literature in SE can be defined as any material about SE that is not formally peer-reviewed nor formally published.''} As GL includes many different types, we advocate using ``shades of grey'' in SE to classify GL material, as proposed by Garousi et al.~\cite{Garousi:2019:IST}, to avoid misunderstanding about its types.}

\vspace{0.2cm}
\noindent
\textbf{Challenge 2: Lack of search efforts for Grey Literature in specific data sources}. Our investigation for RQ3 observed most of the studies using Google (search or scholar) as a primary source to search for studies. To search for GL, Bonato~\cite{Bonato:SearchingGLBook:2019} emphasized the importance of using specialized data sources because they are reproducible, and these sources provide a means to identify Deep Web content, while Google may not identify more than 16\% of the content available. In the SE area, several specialized data sources provide important GL content that could be useful for researchers, for instance, blogs, Q\&A websites, and videos.

\MyBox{\noindent \textit{Potential way(s) to address:} As GL in SE can be published in different ways, it is essential to understand the sources that could provide valuable information to the research and understand the viability to use the data available because each source provides different characteristics. This advice was also partially recommended by Garousi et al.~\cite{Garousi:2019:IST}. Another issue to be considered is how to find relevant and rigorous GL in a considerable amount of information that could be retrieved if deemed the use of search engines (e.g., Google). Rainer and Williams~\cite{Rainer:2019:IST:Heuristic} proposed using heuristics to improve GL searches' relevance and rigor to address this challenge. These recommendations avoid retrieving a vast amount of irrelevant data.}

\vspace{0.2cm}
\noindent
\textbf{Challenge 3: Lack of specific quality assessment criteria for Grey Literature and its specific types}. By analyzing RQ3, we noticed that most of the Secondary Studies that included GL (126) did not employ specific criteria for quality assessment, even amongst the studies that explicitly search for GL. Despite that, some previous studies perceived the difference in GL studies' nature compared to traditional literature~\cite{Garousi:2019:IST}, suggesting that these studies need to be evaluated in different ways. Moreover, Tom et al.~\cite{SS67} reported that due to the heterogeneity of the studies included investigated in an MLR, it was necessary to consider each type of GL's specific nature. This claim is also supported by Garousi et al.~\cite{Garousi:2019:IST}, although it is not restricted to studies that included GL. Kitchenham~\cite{Kitchenham:2007:Guideline} also drew attention to quality assessment instruments that meet the different types of studies.

\MyBox{\noindent \textit{Potential way(s) to address:} When looking for GL, SE researchers should define a set of quality assessment criteria appropriate to assess these studies, in particular, by observing if the requirements are adequate for GL types retrieved in the search. Although some previous studies have already defined some quality criteria assessment for GL (e.g.,~\cite{Garousi:2019:IST,Soldani:2018:GLR}), we believed that more effort and attention by the SE research community is needed.}

\vspace{0.2cm}
\noindent
\textbf{Challenge 4: Lack of Grey Literature classification}. Our analysis for RQ4 found only 46\% of secondary studies 
classifying the GL studies they have used. This lack of classification increased our effort to interpret the GL used (e.g., accessing the online link available), which could also introduce additional interpretation bias. This problem hinders a comprehensive understanding of the types of GL used, for example, to understand better which types of GL are commonly investigated.

\MyBox{\noindent \textit{Potential way(s) to address:} As occurs with scientific papers included in Secondary Studies, they are usually classified by the publication channel (e.g., as a journal, conference, or workshop paper); we highlight the importance of classifying the GL  with their types for the reader to understand what types were used and to guide future research that may want to investigate specific GL types.}

\vspace{0.2cm}
\noindent
\textbf{Challenge 5: Grey Literature availability.} Our analysis for RQ4 investigated 1,246 Grey Literature included in investigated Secondary Studies (n=126). We found that 24.8\% did not provide the GL URLs, and almost half of the URLs informed were not working. Farace and Sch\"{o}\-pfel~\cite{Schopfel:2010:Book} also recognized this problem with GL availability. It happened because some websites were broken or the URLs had changed. This challenge hinders the appraisal of the evidence retrieved and limits the secondary studies' replicability that used Grey Literature.

\MyBox{\noindent \textit{Potential way(s) to address:} For this challenge, we perceived two possibles ways to deal with this challenge. The first one is storing all data searched and collected in an external database (preference for Open Access) for later consultation, such as archiving data on preserved archives such as Zenodo and Figshare, as recommended by Mendez et al.~\cite{Mendez:OpenScience:2020}. Although websites that could significantly mitigate this problem exist, we believe we need a more robust culture to widen searches or promote permanent GL. The second one is trying to minimize this challenge using web archiving initiatives (e.g., Internet Archive\footnote{https://archive.org}) that preserves information published on the web or digitized from printed publications~\cite{Costa:InternetArchive:2017}. For example, accessing one GL URL\footnote{http://weblogs.asp.net/astopford/archive/2010/07/19/technical-debt.aspx} is returned that the page was not found. Nevertheless, by using the Internet Archive, we could found the web content.}

\vspace{0.2cm}
\noindent
\textbf{Challenge 6: Lack of reliability/credibility}. Our investigation for RQ5 and RQ6 found that even with the perceived benefits and motivations to use GL, several researchers avoid using it due to the lack of reliability or credibility~\cite{RQSS19,RQSS02,RQSS22}. This trade-off between the benefits of ``hearing the practitioners voice'' and ``lack of reliability or credibility'' were expected, in part, but they also show the need for further investigation on how to improve the selection of content provided in GL and to better deal with it.

\MyBox{\noindent \textit{Potential way(s) to address:} One possible way to deal with this challenge is selecting GL sources based on the 1st and 2nd tier of the ``shades'' of GL, aiming to retrieve evidence from sources produced by authors with high or moderate expertise and with high or moderate outlet control/credibility of the content production. 
Another possibility is the researchers employed a set of criteria to assess the GL credibility as discussed by previous studies. For example, Kamei et al.~\cite{Kamei:SBES:2020} investigated the importance of selecting sources from renowned authors, institutions, companies, or a renowned producer cited that, and Williams and Rainer~\cite{Williams:2017:EASE} proposed another set of criteria claiming that the GL source needs to be rigorous, relevant, well written, and experience-based. Thus, from these possibilities, SE researchers could take a decision whether GL is suitable to use or not.}



\section{Limitations}\label{sec:limitations}
Like any empirical study, ours also has limitations and threats to validity. 


An internal threat of our research is related to how we interpret and analyze the types of GL used in the selected studies. Such information was not always easy to retrieve, which then forced us to confide in our interpretation. To mitigate this threat, for instance, when a selected study mentioned the amount of GL included, we also manually counted the GL cited in the references to see if we could match the same number. We noticed a negligible difference between what was presented and what we found ($\approx$3\%). However, we noted that our interpretation in classifying GL types into different tiers, sometimes created controversy in relation to certain studies. For instance, we considered books and book chapters as GL, and we noted that some of the studies did not consider them. If we excluded those types of analysis, we would remove 38 out of 126 studies that exclusively used this type of GL. However, we highlight that our interpretation of GL types using the ``shades'' of grey corroborates with previous studies in SE (for instance, the studies of Zhang et al.~\cite{Zhang:2020:ICSE}, Garousi et al.~\cite{Garousi:2019:IST}, and Williams and Rainer~\cite{Williams:2017:EASE}).

We tried to mitigate external validity by selecting a broad set of selected studies published from 2011 up to 2018. These studies are also representative, being published in premier SE conferences and journals. However, our decision to focusing only on top conferences and journals may have introduced a bias of under-represent or over-represent the use of GL. 
Still, although we followed a paired process during this research, we also found some challenges with the list of included secondary studies. For instance, we observed that some SLRs (e.g.,~\cite{SS8,SS51}) and MSs (e.g.,~\cite{SS7,SS88}) did not use the term ``multivocal'', even though they used a systematic search and selection process for GL studies. This may have introduced some bias in our classification because we opted to rely on the authors' classification. Therefore, the number of multivocal research papers found in this work (Figures~\ref{fig:chart-distribution-secondarystudies-years} and \ref{fig:chart-gl-usage}) might underestimate the multivocal studies' overall number. Since the term ``multivocal'' was recently introduced in SE~\cite{Garousi:2016:EASE,garousi2017guidelines,Garousi:2019:IST}, we believe that works' authors might not be aware of this terminology.
For the future, we expect that researchers could adopt the appropriate terminology to mitigate this potential threat. This problem was also previously identified by Garousi and K\"{u}\c{c}\"{u}k~\cite{Garousi:2018:JSS}. Another external threat is related to using a sample of Secondary Studies that did not use GL (n=320) to answer RQ1 and RQ5 because it is possible for the study that did not use GL to have a pre-conceived view of GL. However, comparing the samples of including or not GL, the findings for RQ1 showed the occurrences consistent with each category. Concerning RQ5, it was necessary to use this sample since our investigation explored the motivations to avoid it. Then such information could appear in a study that did not use it.


\section{Conclusions and Future Works}\label{sec:conclusions}

In this paper, we conducted a tertiary study to understand GL's landscape in Secondary Studies of SE. We identified a total of 446 Secondary Studies, within which we investigated 126 (28.2\%) for a more comprehensive understanding.

We found a lack of GL definition among the studies and different interpretations of a GL type. We believe that the use of ``shades'' of grey could be promoted and help solve this challenge.

Our findings have several implications for SE research. We highlighted the importance of GL to Secondary Studies, and we presented several benefits and motivations to use it. We also found some challenges and reasons to avoid GL, showing that future investigations are necessary. Moreover, we discovered the need for specific guidelines to search, select, and assess GL, taking into account GL types' plurality. Researchers should also consider developing methods to improve GL's availability, allowing their data to be preserved and accessed by others in the future. Those guidelines will be important to the SE research to take better advantage of using GL.

By describing our findings and a list of challenges with the potential ways to address them, we expect to help others to use GL in SE research. To conclude, GL is an important source of evidence for Secondary Studies but needs more maturity for researchers' broad acceptance.

For replication purposes, all the data used in this study is available online at: \texttt{\url{https://doi.org/10.5281/zenodo.4079994}}.

\subsection{Future work}

There are many ways in which this work could be complemented, including to understand on a large scale how SE researchers interpret the GL types according to the shades of GL; to investigate possibilities to improve the criteria of quality assessment that attend the different perspective of the nature of the GL types; to investigate approaches to contribute to the maintenance of GL availability. Finally, we also plan to propose a guideline with practices to search, select, and assess GL.

\bibliographystyle{cas-model2-names}

\bibliography{references}




\clearpage
\newpage

\noindent{\large\bfseries Appendix A. Secondary Studies Included\par}
\nocitesec{*}
\bibliographystylesec{appendixStyle}
\bibliographysec{secondaryStudies}

\clearpage
\newpage

\noindent{\large\bfseries Appendix B. Secondary Studies without Grey Literature \par}
\vspace{.3cm}
\noindent Although these studies did not include Grey Literature in their datasets, they had a Grey Literature definition or justified at some level the reasons not to include it on the research, which served as basis to partially answer our research questions RQ1 and RQ5.

\nocitesecNot{*}
\bibliographystylesecNot{appendixStyle}
\bibliographysecNot{secondaryStudiesNotIncluded}

\end{document}